\documentclass[aps,pra,a4paper,twocolumn,10pt]{revtex4-2}
\usepackage{graphicx}
\usepackage{amsmath}
\usepackage{amssymb}
\usepackage{mathrsfs}
\usepackage{amsfonts}
\usepackage{dsfont}
\usepackage{dcolumn}
\usepackage{bm}
\usepackage{booktabs}
\usepackage{color}
\usepackage{xcolor}
\usepackage{colortbl}
\usepackage{mathtools}
\usepackage{upgreek}
\usepackage{cancel}

\providecommand{\abs}[1]{\left|#1\right|}

\providecommand{\ket}[1]{|#1\rangle}
\providecommand{\bra}[1]{\langle#1|}
\providecommand{\brak}[2]{\langle#1|#2\rangle} 
\providecommand{\proj}[2]{|#1\rangle \! \langle#2|} 
\providecommand{\mean}[3]{\langle#1|#2|#3\rangle} 
\newcommand{\di}{\textrm{d}}

\newcommand{\deff}{{\, \vcentcolon = \,}}

\begin{document}

\title{A possible statistics loophole in Bell's theorem}

\author{Andrea Aiello}
\email{andrea.aiello@mpl.mpg.de} 
\affiliation{Max Planck Institute for the Science of Light, Staudtstrasse 2, 91058
Erlangen, Germany}


\date{\today}

\begin{abstract}
Bell's theorem proves the incompatibility between quantum mechanics and local realistic hidden-variable theories. In this paper we  show that, contrary to a common belief, the theoretical proof of Bell's theorem is not affected by  counterfactual reasoning. Then, we demonstrate that the experimental verification of this theorem may be affected in an unknowable way by our ignorance about the probability distribution of the hidden variables. Our study is based on the standard theory of random variables, and lays the groundwork for a critical rethinking of Bell's theorem and its consequences.
\end{abstract}

\maketitle

\section{Introduction}\label{intro}

Quantum mechanics is at the same time the most successful and the most debated theory in physics we have, still today more than one hundreds years after its foundation. 
One of the reasons for this debate originated from the celebrated paper by Einstein, Podolsky and Rosen \cite{PhysRev.47.777}, where the authors discussed the possible incompleteness of quantum mechanics (QM), and the need to complete it by introducing additional variables, which later became known as hidden variables.  
The rest is well-known history and needs not be repeated here. We only point out that the turning point in that history was represented by the so-called Bell's theorem and its by-products \cite{sep-bell-theorem}. In short, Bell proved that no local, realistic hidden-variable (HV) theory  could describe the outcomes of certain  correlation experiments with entangled particle \cite{Bell1964,RevModPhys.38.447}. He was able to express his theorem in the form of an inequality that could be verified experimentally. Since then, many variants of such inequality have been derived \cite{Ballentine}, like the largely used Clauser, Horne, Shimony and Holt  (CHSH) inequality  \cite{PhysRevLett.23.880}.
More recently, Bell's theorem has made a comeback mainly because of its usage in the field of quantum information \cite{wilde_2017}.
Despite these successes, Bell's theorem has not been free from criticisms \cite{DeBaere1984,Bohm1986,ADENIER2001,Cetto2020,Lambare2021,Czachor_2023}, the most significant of which is perhaps the use of counterfactual reasoning \cite{Peres1995,Hance_2024}.  

The purpose of this work is twofold. First, we demonstrate that, in fact, there is no counterfactual reasoning in the proof of CHSH's inequality. Then, we show that any experimental verification of this inequality is inevitably affected by a loophole rooted in our ignorance about the statistical distribution of the hidden variables. 
The tools we use to obtain our results are simply basic quantum mechanics and random variable theory.

This paper is structured as follows. In Sec. \ref{CHSHreview}, we present a quick review of two-photon correlation experiments. Next, in Sec. \ref{deriveCHSH} we show 
that counterfactual reasoning is not necessary to prove the CHSH inequality. This inequality is usually written in terms of the so-called Bell parameter $S$, as $\abs{S} \leq 2$. The parameter  $S$ is given by the mean value of a suitably defined random variable $T$, which describes a specific linear combination of the outcomes of four distinct correlation experiments. In Sec. \ref{BellPar} we perform a detailed statistical analysis of $T$ showing that the \emph{sample mean} of $T$, which is the quantity actually measured in real-world experiments, may differ from $S$, the \emph{mean} value of $T$, in an unknowable way.
Then, in Sec. \ref{measuring} we calculate the actual upper bound of CHSH's inequality. Finally, in Sec. \ref{conclusions}, we draw our conclusions. Four appendixes reporting detailed calculations, complete the paper.

\section{A brief review of two-photon correlation experiments}\label{CHSHreview}

Consider a light source  that emits pairs of counterpropagating photons, say photon $A$ and photon $B$, in the entangled state
\begin{align}\label{a10}
\ket{\psi} & =  \frac{1}{\sqrt{2}} \Bigl(  \ket{e_+^{(A)}} \otimes \ket{e_-^{(B)}} -  \ket{e_-^{(A)}} \otimes \ket{e_+^{(B)}} \Bigr) \nonumber \\[6pt]
& =  \frac{1}{\sqrt{2}} \Bigl(  \ket{e_+^{(A)},e_-^{(B)}} -  \ket{e_-^{(A)},e_+^{(B)}} \Bigr),
\end{align}
where the two vectors $\ket{e_k^{(A)}}, ~ (k = \pm 1)$,  form a complete and orthonormal basis in the two-dimensional Hilbert space $\mathcal{H}_A$ representing the polarization state of photon $A$. The same definitions hold, mutatis mutandis, for photon $B$. By construction, $\ket{\psi} \in \mathcal{H} = \mathcal{H}^{(A)} \otimes \mathcal{H}^{(B)}$, where $\mathcal{H}$ is the four-dimensional Hilbert space of the pair of photons.
Note that throughout this paper, when an index takes on the two values $\pm 1$, we write $q_{\pm}$ as an abbreviation of $q_{\pm 1}$, for the generic quantity $q_k, ~ (k = \pm 1)$.

After being emitted, photons $A$ and $B$ propagate to Alice and Bob, respectively, who can test the linear polarization of the photons they receive, using rotatable polarization analyzers. 
Alice's analyzer has one input port, where the photons enter, and two output ports, labeled $d_{A+}$ and $d_{A-}$, from which the photons can exit.
If a photon exits the analyzer from the port $d_{A+}$, its polarization is described by the vector state 
\begin{align}\label{a12}
\ket{\zeta_+^{(A)}(\alpha)}  = \cos \alpha \ket{e_+^{(A)}} + \sin \alpha \ket{e_-^{(A)}},
\end{align}
 where $\alpha \in [0,\pi)$ is the angle by which Alice's polarization  analyzer is rotated. Conversely, when the photon exits  from the port $d_{A-}$, its polarization is represented by the vector state 
\begin{align}\label{a14}
\ket{\zeta_-^{(A)}(\alpha)}  = -\sin \alpha \ket{e_+^{(A)}} + \cos \alpha \ket{e_-^{(A)}}.
\end{align}
Again, the same definitions apply to photon $B$ replacing everywhere $A$ with $B$ and $\alpha$ with $\beta$. 
Behind each of Alice and Bob's  output ports of their analyzers, there is a detector that fires when reached by a photon.

In this context, the term ``experiment'' denotes the emission of a pair of photons by the light  source, followed by the firing of a pair of Alice's and Bob's detectors.
The possible outcomes of this experiment are given by the set $\Omega$ (sample space of the experiment), whose elements are all the ordered pairs $(d_{Ak},d_{Bl}) \deff \omega_{kl},~ (k,l=\pm 1)$, defined by
\begin{align}\label{a20}
\Omega & = \left\{ (d_{A+}, d_{B+}), (d_{A+}, d_{B-}), \right. \nonumber \\[6pt]
& \phantom{= \left\{ \right.} \left. (d_{A-}, d_{B+}), (d_{A-}, d_{B-}) \right\}\nonumber \\[6pt]
& \deff \left\{ \omega_{++}, \omega_{+-}, \omega_{-+}, \omega_{--} \right\},
\end{align}
where $+$ and $-$ stand for $+1$ and $-1$, respectively. Then, for example,  the outcome $\omega_{+-} = (d_{A+}, d_{B-})$ describes the event where  Alice's detector behind port $d_{A+}$ has fired, and Bob's detector behind $d_{B-}$, has also fired.
We remark that each experiment, which we denote by $E$, is characterized by an ordered pair of angles $(\alpha,\beta)$, which describes the orientation of Alice and Bob's analyzers respectively; hence we write $E = E(\alpha,\beta)$. Different pairs of angles characterize different experiments, so that  $E(\alpha,\beta) \neq E(\alpha',\beta')$, when $(\alpha,\beta) \neq (\alpha',\beta')$.

Next, we define the two dichotomic random variables $X$ and $Y$, as
\begin{align}\label{a30}
X(\omega_{kl}) = k, \qquad \text{and} \qquad Y(\omega_{kl}) = l, 
\end{align}
where $k,l = \pm 1$. By hypothesis, $X$ and $Y$ have respective probabilities
\begin{align}\label{a40}
P(X = k) = p_k, \qquad \text{and} \qquad P(Y = l) = q_l, 
\end{align}
with $0 \leq p_k, q_l \leq 1, ~ (k,l=\pm 1)$.
Their joint probabilities are the four nonnegative numbers
\begin{align}\label{a50}
P(X = k, Y = l) = p_{kl}, 
\end{align}
with
\begin{align}\label{a60}
\sum_{k,l = \pm 1} p_{kl}  = 1,
\end{align}
and
\begin{align}\label{a70}
p_k = \sum_{l = \pm 1} p_{kl}, \qquad  q_l  = \sum_{k = \pm 1} p_{kl}, \qquad (k,l=\pm 1).
\end{align}
By definition, the four probabilities $p_{kl}$ depends on the analyzers' orientation angles $\alpha$ and $\beta$, that is 
\begin{align}\label{a75}
p_{kl} = p_{kl}(\alpha,\beta).
\end{align}
As shown in \cite{PhysRevA.87.022114}, for a pair of dichotomic random variables $X = \pm 1$ and $Y = \pm 1$, it is always possible to write
\begin{multline}\label{a50bis}
P(X = k, Y = l) \\[6pt] = \frac{1}{4}  + k \, \frac{\mathbb{E}[X]}{4} + l \, \frac{\mathbb{E}[Y]}{4} + k  \, l \, \frac{\mathbb{E}[X Y]}{4} ,
\end{multline}
with $ k,l=\pm 1$, and here and hereafter $\mathbb{E}[U]$ denotes the expectation value of the random variable $U$.

Since the vector state \eqref{a10} is invariant with respect to rotations in the polarization space,  whatever theory we use to calculate the probabilities $p_{kl}(\alpha,\beta)$, the marginal probabilities must satisfy $p_+ = p_- = 1/2$ and $q_+ = q_- =1/2$. This implies that the random variables $X$ and $Y$ have zero mean,
\begin{align}\label{a80}
\mathbb{E}[X] = 0 = \mathbb{E}[Y],
\end{align}
and unit variances,
\begin{align}\label{a90}
\mathbb{E}[X^2] = 1 = \mathbb{E}[Y^2] .
\end{align}
Therefore, if we write the  linear correlation coefficient between $X$ and $Y$ as,
\begin{align}\label{a100}
C(\alpha,\beta) \deff \mathbb{E}[X Y] = \sum_{k,l=\pm 1} k l \, p_{kl}(\alpha,\beta),
\end{align}
then from (\ref{a50bis}-\ref{a90}) it follows that 
\begin{align}\label{a105}
p_{kl}(\alpha,\beta) = \frac{1}{4} + k \, l \,\frac{C(\alpha,\beta)}{4} .
\end{align}
This expression is valid irrespective of whatever theory we use to calculate $C(\alpha,\beta)$, either quantum mechanics or a hidden-variable theory. The only  constraint to $C(\alpha,\beta)$ is the normalization
\begin{align}\label{a106}
-1 \leq  C(\alpha,\beta)  \leq 1.
\end{align}
For completeness, we also calculate the variance $\mathbb{V}[X Y]$ of $X Y$ \cite{papoulis2002}, as
\begin{align}\label{a162}
\mathbb{V}[X Y] & = \mathbb{E}[X^2 Y^2] - \left( \mathbb{E}[X Y] \right)^2 \nonumber \\[6pt]
& =  1  - C^2(\alpha,\beta).
\end{align}

The four probabilities $p_{kl}$ can be estimated from the experimental data by repeating the experiment $N \gg 1$ times and using the frequency interpretation of probability \cite{papoulis2002}. Let $N_{kl},~(k,l=\pm 1)$ be the number of simultaneous clicks of the detectors behinds port $d_{Ak}$ and $d_{Bl}$, after $N$ runs of the experiment. Then, we estimate 
\begin{align}\label{a110}
p_k \cong \frac{N_{k+} + N_{k-}}{N}, \qquad q_l \cong \frac{N_{+l} + N_{-l}}{N},
\end{align}
and
\begin{align}\label{a120}
p_{kl} \cong \frac{N_{kl}}{N}, 
\end{align}
with $N_{++} + N_{+-} + N_{-+} + N_{--} = N$. From \eqref{a100} and \eqref{a120} it follows that
\begin{align}\label{a130}
\mathbb{E}[X Y]   \cong \frac{N_{++} - N_{+-} - N_{-+} + N_{--}}{N}. 
\end{align}

Alternatively, the correlation coefficient $\mathbb{E}[X Y]$ can be calculated by using either \emph{a}) quantum mechanics,  or  \emph{b}) some hypothetical hidden-variable theory. Bell's theorem \cite{Ballentine} basically consists of comparing certain linear combinations of correlation coefficients calculated according to  \emph{a})   and \emph{b}). 
A popular form of Bell's theorem is the so-called CHSH's inequality. It is often stated in textbooks that counterfactual reasoning is required to derive this inequality \cite{Peres1995,CoTannoBookIII}. In the next section we will show that this belief is unjustified.

\section{There is not counterfactual reasoning in the derivation of the CHSH inequality}\label{deriveCHSH}

To obtain the CHSH inequality \cite{PhysRevLett.23.880}, we must run four distinct  experiments,  labeled  $E_\mu(\alpha_\mu, \beta_\mu)$, with $\mu=1,2,3,4$, which are identical to the experiment $E(\alpha,\beta)$ described above, apart from the different orientations of Alice and Bob's polarization analyzers. Consequently, we need to define  four distinct pairs of dichotomic random variables, say $(X_1,Y_1), (X_2,Y_2), (X_3,Y_3)$ and $(X_4,Y_4)$, which represent the outcomes of the experiments $E_1, E_2, E_3$ and $E_4$, respectively.

Note that here and hereafter we will use a notation that is a natural generalization of the one previously established. However, to avoid possible ambiguities, we give the details of  this notation in Appendix \ref{Anotation}.

\subsection{General facts}\label{general}

As first step, we calculate the so-called Bell parameter $S=S \bigl(\alpha,   \beta,  \alpha', \beta' \bigr)$, defined by 
\begin{align}\label{a140}
S \bigl(\alpha, & \;  \beta,  \alpha', \beta' \bigr) \nonumber \\[6pt]
& =  \mathbb{E} \left[ X_1 Y_1 \right] +  \mathbb{E} \left[ X_2 Y_2 \right]  + \mathbb{E} \left[ X_3 Y_3 \right] - \mathbb{E} \left[ X_4 Y_4 \right] \nonumber  \\[6pt]
& =   C(\alpha,\beta) + C(\alpha,\beta')  + C(\alpha',\beta)  - C(\alpha',\beta'),
\end{align}
where 
\begin{align}\label{a150}
C(\alpha_\mu,\beta_\mu) \deff \mathbb{E}[X_\mu Y_\mu] = \sum_{k,l=\pm 1} k l \, p_{kl}(\alpha_\mu,\beta_\mu), 
\end{align}
with $\mu=1,\ldots,4$, and
\begin{equation}\label{a160}
\begin{split}
(\alpha_1,\beta_1) & = (\alpha,\beta),   \\[6pt]
(\alpha_2,\beta_2) & = (\alpha,\beta'),  \\[6pt]
(\alpha_3,\beta_3) & = (\alpha',\beta), \\[6pt]
(\alpha_4,\beta_4) & = (\alpha',\beta'). \\[10pt]
\end{split}
\end{equation}
%

The probability distribution $p_{kl}(\alpha_\mu,\beta_\mu)$ of the outcomes of the experiment $E_\mu(\alpha_\mu, \beta_\mu)$ can be calculated either using quantum mechanics, thus obtaining 
 $p_{kl}^\text{QM}(\alpha_\mu,\beta_\mu)$, or local realistic hidden-variable theories which supposedly give $p_{kl}^\text{HV}(\alpha_\mu,\beta_\mu)$. What is important to remark here is that in both cases, since the four experiments $E_1, E_2, E_3$, and $ E_4$ are supposedly performed independently, the joint probability distribution of the eight random variables $X_1, Y_1, X_2,Y_2, X_3,Y_3,X_4,Y_4$, is given by the product of the individual distributions of the outcomes of the four experiments, viz.,
\begin{widetext}
\begin{align}\label{a165}
P(X_1=k_1, Y_1=l_1, \, X_2=k_2, Y_2=l_2, \, X_3=k_3, Y_3=l_3, \, X_4=k_4, Y_4=l_4) = \prod_{\mu=1}^{4} p_{k_\mu l_\mu}(\alpha_\mu,\beta_\mu),
\end{align}
with $k_\mu, l_\mu = \pm 1$.
\end{widetext}

Quantum mechanics and local realistic hidden-variable  theories predict different values for the correlation functions $C(\alpha_\mu,\beta_\mu)$. Per each individual experiment $E_\mu(\alpha_\mu, \beta_\mu)$, quantum mechanics gives (see Appendix \ref{prob} for detailed calculation), 
\begin{align}\label{a170}
C^\text{QM}(\alpha_\mu, \beta_\mu) & = \sum_{k,l=\pm 1} k l \, p_{kl}^\text{QM}(\alpha_\mu,\beta_\mu) \nonumber \\[6pt]
& = -\cos \bigl[ 2( \alpha_\mu - \beta_\mu )\bigr],
\end{align}
while local realistic HV theories yield \cite{Bell1964},
\begin{align}\label{a180}
C^\text{HV}(\alpha_\mu, \beta_\mu) & = \sum_{k,l=\pm 1} k l \, p_{kl}^\text{HV}(\alpha_\mu,\beta_\mu) \nonumber \\[6pt]
& = \int_\mathcal{D} f_\Lambda(\lambda) a(\alpha_\mu,\lambda) b(\beta_\mu,\lambda) \, \di \lambda.
\end{align}
In Eq. \eqref{a180} $f_\Lambda(\lambda) \deff \rho(\lambda)$ denotes the probability density function of the   random vector $\Lambda$ describing the hidden variables,  which has  domain $\mathcal{S}$ and  range $\mathcal{D}$: 
\begin{align}\label{a190}
\Lambda \, : \, \mathcal{S} \rightarrow \mathcal{D},
\end{align}
and is normalized to,
\begin{align}\label{a210}
\int_\mathcal{D} \rho(\lambda) \, \di \lambda = 1.
\end{align}
It is not necessary to specify how many components $\Lambda$ has, nor whether they are discrete, continuous, or mixed. Following the common convention, here and in the following we will write  $\Lambda$ and  $\lambda$ as if they were continuous one-dimensional parameters.
The two functions $a(\alpha_\mu,\lambda)$ and $b(\beta_\mu,\lambda)$, supposedly take only two values, 
\begin{align}\label{a220}
a(\alpha_\mu,\lambda) = \pm1, \qquad b(\beta_\mu,\lambda)  = \pm 1,
\end{align}
and determine the results of the outcomes of the measurements performed by Alice and Bob, respectively. When a photon pair is emitted, the random variable $\Lambda$ takes a specific value $\lambda$, and this determines the either $+1$ or $-1$ values of $a(\alpha_\mu,\lambda)$ and $b(\beta_\mu,\lambda)$. At this point it is convenient to consider the angles $(\alpha, \beta, \alpha', \beta')$ given and fixed, so that we can simplify the notation renaming 
\begin{equation}\label{a222}
\begin{split}
a(\alpha,\lambda) & = a(\lambda),    \\[6pt]
a(\alpha',\lambda) & = a'(\lambda),  \\[6pt]
b(\beta,\lambda) & = b(\lambda),     \\[6pt]
b(\beta',\lambda) & = b'(\lambda),
\end{split}
\end{equation}
for all $\lambda \in \mathcal{D}$.  These functions are what mathematicians call \emph{simple functions} \cite{RudinPrinciples}, that is measurable functions that takes only finitely many values. They can be written as,
\begin{equation}\label{a223}
\begin{split}
a(\lambda) & = \sum_{k = \pm 1} k \, \chi_{\mathcal{A}_k}(\lambda),   \\[6pt]
a'(\lambda) & = \sum_{k = \pm 1} k \, \chi_{\mathcal{A}_k'}(\lambda),  \\[6pt]
b(\lambda) & = \sum_{l = \pm 1} l \, \chi_{\mathcal{B}_l}(\lambda),   \\[6pt]
b'(\lambda) & = \sum_{l = \pm 1} l \, \chi_{\mathcal{B}_l'}(\lambda),
\end{split}
\end{equation}
where $\chi_{\mathcal{F}}(x)$ denotes the indicator function of the set $\mathcal{F}$, defined by
\begin{align}\label{a224}
\chi_{\mathcal{F}}(x) \deff \left\{
                              \begin{array}{ll}
                                1, & \text{if} \; x \in \mathcal{F} , \\[6pt]
                                0, & \text{if} \; x \not\in \mathcal{F}.
                              \end{array}
                            \right.
\end{align}
Thus, for example, if $\lambda \in \mathcal{A}_+ $, then $a(\lambda) = +1$.
Note that by definition
%
%
%
%
%
%
%
%
\begin{equation}\label{a223bis}
\begin{split}
\mathcal{A}_+ \cup \mathcal{A}_- & = \mathcal{D} = \mathcal{A}_+' \cup \mathcal{A}_-',   \\[6pt]
\mathcal{B}_+ \cup \mathcal{B}_- & = \mathcal{D} = \mathcal{B}_+' \cup \mathcal{B}_-'. \\[10pt]
\end{split}
\end{equation}
Then, from \eqref{a180} and \eqref{a223}, it follows that
\begin{align}\label{a225}
C^\text{HV}(\alpha,\beta) & = \int_\mathcal{D} \rho(\lambda) a(\alpha,\lambda) b(\beta,\lambda) \, \di \lambda \nonumber \\[6pt]
& = \sum_{k,l= \pm 1} k l \int_\mathcal{D} \rho(\lambda) \chi_{\mathcal{A}_k}(\lambda) \chi_{\mathcal{B}_l}(\lambda) \, \di \lambda \nonumber \\[6pt]
& = \sum_{k,l= \pm 1} k l \int_{\mathcal{A}_k \cap \mathcal{B}_l} \rho(\lambda) \, \di \lambda,
\end{align}
where we have used the following property of the indicator functions: $\chi_{\mathcal{F}}(x) \chi_{\mathcal{G}}(x) = \chi_{\mathcal{F} \cap \mathcal{G}}(x)$. Comparing \eqref{a225} with the first line of \eqref{a180}, we infer that 
\begin{equation}\label{a226}
\begin{split}
p_{kl}^\text{HV}(\alpha,\beta) & = \int_{\mathcal{A}_k \cap \mathcal{B}_l} \rho(\lambda) \, \di \lambda,   \\[6pt]
p_{kl}^\text{HV}(\alpha,\beta') & = \int_{\mathcal{A}_k \cap \mathcal{B}_l'} \rho(\lambda) \, \di \lambda,  \\[6pt]
p_{kl}^\text{HV}(\alpha',\beta) & = \int_{\mathcal{A}_k' \cap \mathcal{B}_l} \rho(\lambda) \, \di \lambda,   \\[6pt]
p_{kl}^\text{HV}(\alpha',\beta') & = \int_{\mathcal{A}_k' \cap \mathcal{B}_l'} \rho(\lambda) \, \di \lambda.
\end{split}
\end{equation}

\subsection{The CHSH inequality}\label{inequality}

Each distinct experiment $E_\mu(\alpha_\mu,\beta_\mu)$ is characterized by a different random vector $\Lambda_\mu$, with $\mu=1,2,3,4$, and these four random vectors are \emph{independent and identically distributed} according to the same probability density function $\rho(\lambda)$. Therefore, the joint probability density function of $\Lambda_1 ,\Lambda_2 ,\Lambda_3 ,\Lambda_4$ is given by 
\begin{align}\label{a230}
f_{\Lambda_1 \Lambda_2 \Lambda_3 \Lambda_4}(\lambda_1,\lambda_2,\lambda_3,\lambda_4) & = \prod_{\mu=1}^4f_{\Lambda_\mu}(\lambda_\mu) \nonumber \\[6pt]
& = \prod_{\mu=1}^4 \rho(\lambda_\mu).
\end{align}
Then, using \eqref{a180} and \eqref{a230}, we can rewrite \eqref{a140} as,
\begin{widetext}
\begin{align}
S \left(\alpha,  \beta,  \alpha', \beta' \right) = & \;
    C(\alpha,\beta) + C(\alpha,\beta')  + C(\alpha',\beta)  - C(\alpha',\beta') \nonumber \\[6pt]
= & \; \int_{\mathcal{D}^4}a(\alpha,\lambda_1) b(\beta,\lambda_1) \prod_{\mu=1}^4 f_{\Lambda_\mu}(\lambda_\mu) \, \di \lambda_\mu
+ \int_{\mathcal{D}^4}a(\alpha,\lambda_2) b(\beta',\lambda_2) \prod_{\mu=1}^4 f_{\Lambda_\mu}(\lambda_\mu) \, \di \lambda_\mu
\nonumber \\[6pt]
&  +  \int_{\mathcal{D}^4}a(\alpha',\lambda_3) b(\beta,\lambda_3) \prod_{\mu=1}^4 f_{\Lambda_\mu}(\lambda_\mu) \, \di \lambda_\mu
- \int_{\mathcal{D}^4}a(\alpha',\lambda_4) b(\beta',\lambda_4) \prod_{\mu=1}^4 f_{\Lambda_\mu}(\lambda_\mu) \, \di \lambda_\mu,  \label{a232} \\[6pt]
= & \; \int_{\mathcal{D}}a(\lambda_1) b(\lambda_1)  \rho(\lambda_1) \, \di \lambda_1
+ \int_{\mathcal{D}}a(\lambda_2) b'(\lambda_2) \rho(\lambda_2) \, \di \lambda_2
\nonumber \\[6pt]
&  +  \int_{\mathcal{D}}a'(\lambda_3) b(\lambda_3)  \rho(\lambda_3) \, \di \lambda_3
- \int_{\mathcal{D}}a'(\lambda_4) b'(\lambda_4)  \rho(\lambda_4) \, \di \lambda_4, \label{a235} \\[6pt]
= & \; \int_{\mathcal{D}} \rho(\lambda) \Bigl[ a(\lambda) b(\lambda) + a(\lambda) b'(\lambda) + a'(\lambda) b(\lambda) - a'(\lambda) b'(\lambda) \Bigr] \di \lambda. \label{a240}
\end{align}
\end{widetext}
There is a reason why we wrote this equation in three fully detailed steps, as we will soon see. 

It is easy to show that the quantity
\begin{align}\label{a250}
M(\lambda) & = a(\lambda) b(\lambda) + a(\lambda) b'(\lambda) \nonumber \\[6pt]
& \phantom{=} + a'(\lambda) b(\lambda) - a'(\lambda) b'(\lambda),
\end{align}
satisfies
\begin{align}\label{a260}
M(\lambda) = \pm 2.
\end{align}
To this end, it suffice to rewrite $M(\lambda)$ as,
\begin{align}\label{a270}
M(\lambda) = (a + a')b + (a - a') b'.
\end{align}
Clearly, if $a + a' =0$, then $a - a' = \pm 2$; and if $a -  a'=0$,  then $a + a' =\pm 2$. This implies $M(\lambda) = \pm 2$. Therefore, from \eqref{a240} it follows that
\begin{align}\label{a280}
\abs{S \left(\alpha,  \beta,  \alpha', \beta' \right)} \leq 
 \int_{\mathcal{D}} \rho(\lambda) \abs{M(\lambda)} \leq 2.
\end{align}
This is the celebrated CHSH inequality. It is not difficult to see that this demonstration remains perfectly valid if we replace the four functions $a,a',b,b'$, with their expressions \eqref{a223}, so that the concerns expressed by Cetto and coworkers \cite{Cetto2020}, about the partitions \eqref{a223bis} of the range  $\mathcal{D}$ of the random parameters $\Lambda$, are not a problem here. 

It has been argued that the quantity $M(\lambda)$ actually does not exist and that counterfactual reasoning is involved. For example, Peres wrote:
\begin{quote}
``\textit{There is no doubt that counterfactual reasoning is involved: the four numbers $a,b,c,d$,} [$a,b,a',b'$ in this work] \textit{cannot be simultaneously known. The first observer can measure either $a$ or $c$, but not both; the second one--either $b$ or $d$. Therefore Eq. (6.29)} [Eq. \eqref{a260} in this work] \textit{involves at least two numbers which do not correspond to any tangible data, and it cannot be experimentally verified.}''
\end{quote}
(Quoted  and adapted from section \textbf{6-4} in Ref. \cite{Peres1995}).
Similarly, Cohen-Tannoudji and coauthors, wrote that,  
\begin{quote}
``[...] \textit{one must then consider that it is meaningless to attribute four well defined values $A$, $A'$, $B$, $B'$} [$a,a',b,b'$ in this work] \textit{to each pair. Since only  a maximum of two of them can be measured in a given experiment, we should not be able to talk about these four numbers or argue about them even as unknown quantities.}''
\end{quote}
(Quoted  and adapted from section \textbf{F-3-b} in Ref. \cite{CoTannoBookIII}).

Despite these latter two statements, it is clear that in our detailed derivation \eqref{a240} we did not make use of quantities that could not be measured. We started from the results of four \emph{independent} experiments in \eqref{a232}, and derived the CHSH inequality from there. Therefore, in our calculations we have not exploited counterfactual reasonings. So, where does the contradiction arise from? The reason underlying this conflict is the linearity of the expectation value of a random variable \cite{ExpectationValue} or, equivalently in this case, the linearity of integration \cite{LiebLoss}. In our derivation we used such linearity to write $S = S \bigl(\alpha,   \beta,  \alpha', \beta' \bigr)$, as
\begin{align}\label{a295}
S & =  \int\limits_{\mathcal{D}^4} a b +  \int\limits_{\mathcal{D}^4} a b'  + \int\limits_{\mathcal{D}^4} a' b - \int\limits_{\mathcal{D}^4} a' b' \nonumber  \\[6pt]
& =  \int\limits_{\mathcal{D}} a b +  \int\limits_{\mathcal{D}} a b'  + \int\limits_{\mathcal{D}} a' b - \int\limits_{\mathcal{D}} a' b' \nonumber  \\[6pt]
& =  \int\limits_{\mathcal{D}} \bigl( a b + a b' +a'b - a' b' \bigr) .
\end{align}
From this equation it is clear that the quantity $ a b + a b' +a'b - a' b'$ does not need to represent a physically measurable quantity, because is just the result of the formal mathematical manipulation that we performed to pass from \eqref{a235} to \eqref{a240}.
 However, in the traditional derivations of the CHSH inequality presented in many textbooks and articles, such linearity is used in reversed manner, that is they first calculate $M(\lambda) = a(\lambda) b(\lambda) + a(\lambda) b'(\lambda) + a'(\lambda)b(\lambda) - a'(\lambda) b'(\lambda)$, which is \emph{physically not measurable}, and then evaluate the average
\begin{align}\label{a310}
S & =  \int\limits_{\mathcal{D}} \bigl( a b + a b' +a'b - a' b' \bigr) \nonumber  \\[6pt]
& =  \int\limits_{\mathcal{D}} a b +  \int\limits_{\mathcal{D}} a b'  + \int\limits_{\mathcal{D}} a' b - \int\limits_{\mathcal{D}} a' b'.
\end{align}
Note the different domain of integration in the first line of \eqref{a295} and the last line of \eqref{a310}.
We have thus demonstrated that contrary to common belief, there is no need to use counterfactual reasoning to derive the CHSH inequality. This is our first main result.

We would like to point out that the linearity discussed above was questioned by Czachor, who showed that some models of HV theories could violate it \cite{Czachor_2023}. Moreover, a proposal has recently been made to measure $M(\lambda)$ on a single entangled photon pair \cite{Virzi_2024}.

\section{The Bell parameter as a random variable}\label{BellPar}

The Bell parameter $S$ defined in \eqref{a140} is a linear combination of average values of product of random variables.  Because of the  linearity of the expectation value of a random variable discussed in the previous section, 
we can also write the Bell parameter $S(\alpha, \beta, \alpha', \beta')$ in \eqref{a140}, as the expectation value of the single random variable $T$, defined by
\begin{align}\label{a166}
T = X_1 Y_1  +   X_2 Y_2 + X_3 Y_3 - X_4 Y_4.
\end{align}
Explicitly, we have
\begin{align}\label{a140bis}
\mathbb{E} \left[ T \right] &  =  \mathbb{E} \left[ X_1 Y_1  +   X_2 Y_2 + X_3 Y_3 - X_4 Y_4 \right] \nonumber \\[6pt]
& =  \mathbb{E} \left[ X_1 Y_1 \right] +  \mathbb{E} \left[ X_2 Y_2 \right]  + \mathbb{E} \left[ X_3 Y_3 \right] - \mathbb{E} \left[ X_4 Y_4 \right] \nonumber  \\[6pt]
& =  S \bigl(\alpha,  \beta,  \alpha', \beta' \bigr) .
\end{align}
From the definition \eqref{a166} and  $X_\mu = \pm 1, \, Y_\mu = \pm 1$, $(\mu = 1,2,3,4)$, it follows that the random variable $T$ may take the five values $\{t_i \}$, $(i=1, \ldots, 5)$, defined by
%
%
%
%
%
\begin{align}\label{a167}
\{t_1,t_2,t_3,t_4,t_5 \} = \{-4, -2, 0, 2, 4 \}.
\end{align}

We can find the probability density function $f_T(t)$ of $T$, using the random variable transformation theorem \cite{Gillespie1983}. A straightforward calculation gives
\begin{align}\label{probT10}
f_T(t)  = \sum_{i=1}^5 \, p_{i} \, \delta (t-t_i),
\end{align}
where the probabilities $p_i = p_i(\alpha, \beta, \alpha', \beta')$ are defined by
\begin{widetext}
\begin{align}\label{probT20}
p_i = \sum_{k_1,l_1 = \pm 1} \sum_{k_2,l_2 = \pm 1} \sum_{k_3,l_3 = \pm 1} \sum_{k_4,l_4 = \pm 1}  p_{k_1 l_1}(\alpha, \beta)
p_{k_2 l_2}(\alpha, \beta') p_{k_3 l_3}(\alpha', \beta)p_{k_4 l_4}(\alpha', \beta')
\delta (t_i, k_1 l_1 + k_2 l_2 + k_3 l_3 - k_4 l_4)   ,
\end{align}
\end{widetext}
with $i=1, \ldots, 5$, and $ \delta(i,j)$ denotes the Kronecker delta function 
\begin{align}\label{probT30}
\delta(i,j) = \begin{cases}
0 &\text{if } i \neq j,   \\
1 &\text{if } i=j.   \end{cases}
\end{align}
In Eq. \eqref{probT20} the probabilities  $p_{k_1 l_1} (\alpha, \beta), \ldots, p_{k_4 l_4} (\alpha', \beta')$ are given by \eqref{a105}. The explicit values of the probabilities $p_i$ are given in Appendix \ref{pn}.  It is instructive to verify that 
\begin{align}\label{a140ter}
\mu & \deff \mathbb{E} \left[ T\right]  \nonumber  \\[6pt]
& = \int_\mathbb{R} t \, f_T(t) \, \di t \nonumber \\[6pt]
& =  \sum_{i=1}^5 t_i \, p_{i}  \nonumber  \\[6pt]
& =   C(\alpha,\beta) + C(\alpha,\beta')  + C(\alpha',\beta)  - C(\alpha',\beta').
\end{align}
Moreover, we can calculate the variance $\mathbb{V}[T]$ of $T$, obtaining
\begin{align}\label{probT40}
\sigma^2 & \deff \mathbb{V} \left[ T\right]  \nonumber  \\[6pt]
& = \mathbb{E} \bigl[T^2 \bigr] - \bigl(\mathbb{E} \bigl[ T \bigr] \bigr)^2  \nonumber  \\[6pt]
&  = \int_\mathbb{R} t^2 \, f_T(t) \, \di t\nonumber - \left[ \int_\mathbb{R} t \, f_T(t) \, \di t \right]^2 \nonumber \\[6pt]
&  =  \sum_{i=1}^5 t_i^2 \, p_{i}  - \left[ \sum_{i=1}^5 t_i \, p_{i} \right]^2 \nonumber \\[6pt]
& =  4 - C^2(\alpha,\beta)- C^2(\alpha,\beta')   \nonumber  \\[6pt]
& \phantom{= .} - C^2(\alpha',\beta)  - C^2(\alpha',\beta'),
\end{align}
in agreement with \eqref{a162}.
We remark that these expressions for $\mathbb{E}[t]$ and $\mathbb{V}[T]$ are valid for any correlation function $C(\alpha_\mu,\beta_\mu)$, either classical or quantum mechanical.

Now, suppose to run the series of four experiments  $\{ E_\mu\} =\{ E_1, \, E_2, \, E_3, \, E_4\}$  $N$ of times. In this case we can estimate the mean $\mathbb{E} \bigl[ T \bigr]$ and the variance $\mathbb{V} \bigl[ T \bigr]$ of $T$, by sampling $T$  $N$  times instead of performing the integrations or the sums in \eqref{a140ter} and \eqref{probT40}. Let $t(n)$ be the value of $T$ obtained in the $n\text{th}$ run of the series of four experiments $\{ E_\mu\}$. Then, the sample mean
\begin{align}\label{p10}
\mu_N = \frac{1}{N} \sum_{n=1}^N t(n),
\end{align}
and the adjusted sample variance
\begin{align}\label{p20}
\sigma_N^2 = \frac{1}{N -1 } \sum_{n=1}^N \bigl[ t(n) - \mu_N \bigr]^2,
\end{align}
are unbiased estimators of $\mathbb{E}[T]$ and $\mathbb{V}[T]$, respectively \cite{taboga2017}. 

Thus, $\mu_N$ is simply the $N$-sample estimation of the Bell parameter $S = \mu = \mathbb{E}[T]$ defined by \eqref{a140}. Equation \eqref{a280} tells us that $\abs{S} \leq 2$ according to HV theories, and from the law of large numbers \cite{FellerI}, we know that $\mu_N \to \mu$ for $N$ sufficiently large. Therefore, in such a limit, we expect that, approximately,  $\abs{\mu_N} \leq 2$. However, since all real-world experiments can be repeated only a \emph{finite} number $N$ of times, the key question is: how large must be $N$ to guarantee the validity of $\abs{\mu_N} \leq 2$? 
In other words, how likely would be to find $\abs{\mu_N} > 2$ for any finite $N$, in the case that the physics of the four experiments $\{E_\mu \}$ were correctly described by a hidden-variable theory?
In the remainder we are going to show that it is impossible to answer this question, due to our ignorance about the random hidden variables $\lambda$, and their distribution $\rho(\lambda)$.

\subsection{Calculating the probability distribution of $\mu_N$}\label{mu}

We want to determine the probability distribution of  $\mu_N$. In cases like this, it is standard procedure to associate a random variable $T_n$, with $n=1, \ldots, N$, to the value $t(n)$ obtained in the $n\text{th}$ series of the four experiments $\{ E_1, \, E_2, \, E_3, \, E_4\}$. By definition, the $N$ random variables $T_1, \ldots, T_N$ are independent and identically distributed (iid), according to
\begin{align}\label{p30}
 T_n \sim f_{T_n}(t) = f_{T_{n'}}(t)  = f_{T}(t)  , \; \;  \forall \, n, n' = 1, \ldots, N,
\end{align}
where $f_{T}(t)$ is given by \eqref{probT10}.

Next, we define the additional random variable $M_N$, as
\begin{align}\label{p40}
M_N \deff \frac{1}{N} \sum_{n=1}^N \, T_n.
\end{align}
From this definition it follows that $\mu_N$ is a value taken by $M_N$ in a run of the series of four experiments $\{ E_\mu \}$.
Using again the random variable transformation theorem \cite{Gillespie1983}, we can calculate the probability density function $f_{M_N}(m)$ of $M_N$. A straightforward calculation gives 
\begin{align}\label{p50}
f_{M_N}(m) = \sum_{i_1=1}^5 \cdots \sum_{i_N=1}^5  p_{i_1} \cdots p_{i_N} \delta \left( m - \frac{1}{N} \sum_{n=1}^N t_{i_n}\right),
\end{align}
where $t_{i_n}$ is defined by \eqref{a167}. From this equation it follows that the probability to find, after performing $N$ times the series  of four experiments $\{ E_\mu \}$, $\mu_N$ bigger than a certain value, say  $ m_0 >0 $, is given by
\begin{align}\label{p60}
P (\mu_N & \geq m_0) \nonumber \\[6pt]
& = \int_{m_0}^\infty f_{M_N}(m) \, \di m \nonumber \\[6pt]
& = \sum_{i_1=1}^5 \cdots \sum_{i_N=1}^5  p_{i_1} \cdots p_{i_N} H \left(  \frac{1}{N} \sum_{n=1}^N t_{i_n} - m_0 \right),
\end{align}
where $H(x)$ denotes the Heaviside step function \cite{Hfunction}.

Now, imagine that we want to test the validity of QM against HV theories by measuring Bell's parameter $S$, choosing a set of angles $\{\alpha, \beta, \alpha', \beta' \}$,  that maximizes the violation of CHSH's inequality. Suppose that, after $N$ repetitions of the four  experiments $\{ E_\mu \}$, we found $\mu_N \approx 2 \sqrt{2}$. Is this result sufficient to rule out the validity of HV theories? To verify this, we should calculate the probability that $\mu_N$ is greater than $2 \sqrt{2}$ when $\mu_N$ and $\sigma_N$ are calculated using a HV theory. The key problem is that we cannot perform such calculation because we are not knowledgeable about the probability distribution of the hidden variables $\lambda$. Unfortunately, using HV theories we can calculate only the mean value $\mu = S$, but not the $N$-sample mean $\mu_N$.  This is the second main result of this work.

In principle, one might think of circumventing this problem by using the central limit theorem \cite{taboga2017}. 
For $N$ large enough, the central limit theorem guarantees that $M_N$ becomes normally distributed, that is
\begin{align}\label{p70}
M_N \sim \text{N}(\mu, \sigma^2/N),
\end{align}
where $\text{N}(\mu, \sigma^2/N)$ denotes the  normal distribution with mean $\mu$ and variance $\sigma^2/N$, and the latter two parameters are calculated using \eqref{a140ter} and \eqref{probT40}, respectively. In this limit of large $N$, the probability to find $\mu_N \geq m_0 >0 $,  is approximately  given by
\begin{align}\label{p75}
P (\mu_N & \geq m_0) \approx \frac{1}{2}  \operatorname{erfc}\left(\frac{m_0 -\mu }{\sqrt{2} \, \sigma/\sqrt{N} }\right),
\end{align}
where $\operatorname{erfc}(z)$ denotes the complementary error function \cite{Erfc}.

Equations \eqref{p70} and \eqref{p70} are remarkable in that they are independent from the values of $p_i$ and, consequently, are valid for both QM and HV theories. However, unfortunately the central limit theorem does not tell us how large $N$ must be to make these equations practically useful \cite{Gillespie1983}. 
The rate of approach of $f_{M_N}(m)$ to $\text{N}(\mu, \sigma^2/N)$ as $N$ increases, depends on the given form of  $f_{M_N}(m)$, that is, on the form of the probabilities $p_i$, which are unknown for HV theories. Clearly, some forms of $p_i$ produce faster convergence than others. As we do not know the random parameters $\lambda$ and their distribution $\rho(\lambda)$, we cannot quantify such a rate of convergence for arbitrary HV theories. So, in practice, the central limit theorem does not help. An elementary example will illustrate the model dependence of the rate of approach of $f_{M_N}(m)$ to $\text{N}(\mu, \sigma^2/N)$.

\subsection{A simple example}\label{simple}

In this example we will test QM against a concrete model of HV theory. Specifically, we choose the simple model proposed by Aspect in  \cite{Aspect1982}. In such a model, 
\begin{align}\label{p80}
a(\theta,\lambda) = \left\{
                                          \begin{array}{ll}
                                            +1, & \; \text{if} \; \; \cos[2(\theta - \lambda)] \geq 0, \\[6pt]
                                            -1, & \; \text{if} \; \; \cos[2(\theta - \lambda)] < 0,
                                          \end{array}
                                        \right.
\end{align}
and $b(\theta,\lambda) = - a(\theta,\lambda)$, with $\theta \in [0,\pi)$, and $\lambda \in \mathbb{R}$ uniformly distributed between $\lambda = 0$ and  $\lambda = \pi$, that is
\begin{align}\label{p90}
\rho(\lambda) = \frac{1}{\pi}, \qquad \lambda \in [0,\pi].
\end{align}
This choice yields the correlation function 
\begin{align}\label{p100}
C^\text{HV}(\alpha,\beta) & = \frac{1}{\pi}  \int_0^\pi a(\alpha,\lambda) b(\beta,\lambda) \, \di \lambda \nonumber \\[6pt]
& = \left\{
\begin{array}{cl}
 3+ \frac{4 (\alpha - \beta) }{\pi }, & -\pi <\alpha - \beta \leq -\frac{\pi }{2}, \\[6pt]
 -1-\frac{4 (\alpha - \beta)}{\pi }, & -\frac{\pi }{2}< \alpha - \beta <0, \\[6pt]
 -1+ \frac{4 (\alpha - \beta) }{\pi }, & 0\leq \alpha - \beta <\frac{\pi }{2}, \\[6pt]
 3-\frac{4 (\alpha - \beta) }{\pi }, & \frac{\pi }{2}\leq \alpha - \beta <\pi. 
\end{array} \right.
\end{align}
Next, we choose the set of angles $\{\alpha, \beta, \alpha', \beta' \}$,  such that the violation of CHSH's inequality is maximal,
\begin{align}\label{p110}
\{\alpha, \beta, \alpha', \beta' \} = \{ -\pi/8, \, \pi/2, \, \pi/8, \, \pi/4 \}.
\end{align}
Then, using \eqref{a140ter} and \eqref{probT40} we find,
\begin{align}\label{p120}
\mu^\text{QM} = 2 \sqrt{2}, \qquad \sigma^\text{QM} = \sqrt{2},
\end{align}
and
\begin{align}\label{p130}
\mu^\text{HV} = 2  \qquad \sigma^\text{HV} = \sqrt{3}.
\end{align}

Let us choose $N=10$, and calculate the probability that $\mu_N^\text{QM} \geq 3 > \mu^\text{QM}$ using both the exact formula \eqref{p60}, and the approximate one \eqref{p75}. A straightforward calculation gives
\begin{align}\label{p140}
P (\mu_{N}^\text{QM} \geq 3 ) \cong \left\{
\begin{array}{cl}
37 \%, & \text{exact}, \\[6pt]
35 \%, & \text{approximate}. 
\end{array} \right.
\end{align}
This shows that for the QM model the value $N=10$ is already good enough to yield a fair approximation. 

Now let us  calculate in the same way the probability that $\mu_N^\text{HV} \geq 2 \sqrt{2} > \mu^\text{HV}$, again with $N=10$. In this case we find,
\begin{align}\label{p150}
P (\mu_{N}^\text{HV} \geq 2 \sqrt{2} ) \cong \left\{
\begin{array}{cl}
4.3 \%, & \text{exact}, \\[6pt]
6.5 \%, & \text{approximate}.
\end{array} \right.
\end{align}
This equation shows that the central-limit-theorem approximation evaluated at $N=10$, is not so good for the HV model and a bigger value of $N$ would be desirable. So, this example shows that  the rate of convergence of $f_{M_N}(m)$ to $\text{N}(\mu, \sigma^2/N)$ as $N$ increases, is higher in the QM model than in the HV model.

We remark that in this example the QM and HV theories were treated equally, in the sense that in both cases we knew the probability distributions of both $\mu_N^\text{QM}$ and $\mu_N^\text{HV}$. However, this is not the case in the actual experiments, where we ignore the details of the HV theories. In essence, it is in this disparity of statistical knowledge between quantum and hidden variables that is rooted the problem of verifying Bell's theorem studied in this work.

\section{Actual bounds on the sample mean of the Bell parameter $S$}\label{measuring}

In the previous section we have seen that HV theories permit to calculate the mean value $S = \mu$ of the random variable $T$, but not the sample mean $\mu_N$. Hidden variables theories also furnish the bounds $-2 \leq S \leq 2$. Therefore, it is natural to wonder what we can say about the bounds of $\mu_N$. That is what we will do in this section.

Let us consider the individual experiment $E_\mu(\alpha_\mu,\beta_\mu)$, with $\mu=1,2,3,4$. To perform it, Alice and Bob rotate their polarization analyzers along the angles $\alpha_\mu$ and $\beta_\mu$, respectively, and repeat the experiment a large number $N$ of times. 
According to HV theories, at each  photon pair emission, the random vector $\Lambda_\mu$ defined in Sec. \ref{deriveCHSH}, takes a specific but random value, say $ \lambda_{\mu,n},~(n=1,\ldots,N) $, so that the $N$ repetitions of the experiment $E_\mu(\alpha_\mu,\beta_\mu)$ are characterized by the list of  $N$ values 
\begin{align}\label{b10}
 \Lambda_\mu = \{ \lambda_{\mu,1}, \lambda_{\mu,2}, \ldots, \lambda_{\mu,N} \},
\end{align}
taken by $\Lambda_\mu$. Let $N_{\mu,kl},~(k,l=\pm 1)$ be the number of simultaneous clicks of Alice and Bob's detectors behinds port $d_{A_\mu k}$ and $d_{B_\mu l}$, respectively.
Then, at the end of the day Alice and Bob measure the following correlation between the outcomes of their measurements,
\begin{align}\label{b20}
\mathbb{E}[X_\mu Y_\mu]   \cong \frac{N_{_\mu,++} - N_{_\mu,+-} - N_{_\mu,-+} + N_{_\mu,--}}{N}. 
\end{align}
This  correlation would be equal, if a HV theory were valid, to
\begin{align}\label{b30}
C^\text{HV}_N(\alpha_\mu, \beta_\mu) = \frac{1}{N} \sum_{n=1}^N a(\alpha_\mu,\lambda_{\mu,n}) b(\beta_\mu,\lambda_{\mu,n}),
\end{align}
where the subscript $N$ reminds that this is a finite version of \eqref{a180}. In more technical terms, $C^\text{HV}_N(\alpha_\mu, \beta_\mu)$ is the sample covariance of the random variables $a(\alpha_\mu,\lambda_{\mu,n})$ and $b(\beta_\mu,\lambda_{\mu,n})$ \cite{papoulis2002}.

Now, suppose to have performed $N$ times the series of  four experiments $\{E_1, E_2, E_3, E_4\}$, thus obtaining the estimate $S_N = \mu_N^\text{HV}$, of the value of the Bell parameter $S$. According to HV theories, $S_N$ would be given by \cite{ADENIER2001},
\begin{align}\label{b40}
S_N & = C^\text{HV}_N(\alpha, \beta) + C^\text{HV}_N(\alpha, \beta') + C^\text{HV}_N(\alpha', \beta) - C^\text{HV}_N(\alpha', \beta) \nonumber \\[4pt]
& = \frac{1}{N} \sum_{n=1}^N \Bigl[ 
a(\lambda_{1,n}) b(\lambda_{1,n}) + a(\lambda_{2,n}) b'(\lambda_{2,n})  \nonumber \\[0pt]
& \phantom{\frac{1}{N} aaaaaa} + a'(\lambda_{3,n}) b(\lambda_{3,n}) -  a'(\lambda_{4,n}) b'(\lambda_{4,n})  \Bigr]\nonumber \\[0pt]
&   \deff \frac{1}{N} \sum_{n=1}^N  M_n,
\end{align}
where we have defined $M_n = M_n(\lambda_{1,n},\lambda_{2,n},\lambda_{3,n},\lambda_{4,n})$, as
\begin{align}\label{b50}
M_n & \deff a(\lambda_{1,n}) b(\lambda_{1,n}) +
a(\lambda_{2,n}) b'(\lambda_{2,n}) \nonumber \\[6pt]
& \phantom{\deff} + a'(\lambda_{3,n}) b(\lambda_{3,n}) -
a'(\lambda_{4,n}) b'(\lambda_{4,n}),
\end{align}
The difference between $M_n$ in \eqref{b50}  and $M(\lambda)$ in \eqref{a250}, is that while $M(\lambda)$ depends on a \emph{single} set of values  $\lambda$, $M_n$ depends on the \emph{four} set of values $\lambda_{1,n},\lambda_{2,n},\lambda_{3,n}$ and $\lambda_{4,n}$. Since the latter are independent numbers sampled from the same probability distribution $\rho(\lambda)$, we have that
\begin{align}\label{b60}
\min M_n = -4, \qquad \text{and} \qquad \max M_n  = 4.
\end{align}
Therefore, one could be tempted to state that $\abs{S_N} \leq 4$, thus contradicting $\abs{S}\leq 2$ in \eqref{a280}. However, this conclusion would not be correct, as we are going to show now.

\subsection{Actual upper bound of  $S_N$}\label{bounds}

Let us analyze the problem from a practical point of view, considering only the measured data at our disposal  \cite{PhysRevLett.49.91,PhysRevLett.81.3563,PhysRevLett.81.5039,MandelBook,sep-bell-theorem}.
To measure $S_N$ we must perform $N$ times the series of four experiments $\{ E_\mu \}$, that is we must generate $4N$ distinct photon pairs. After we had ran the four experiments $N$ times, at the end of the day we got an array of $4N$ values, say $V_N$, given by
\begin{align}\label{b65}
V_N = \frac{1}{N} \begin{bmatrix}
 \vphantom{\Bigl[} 
  x_{1,1}y_{1,1} & x_{2,1}y_{2,1} & x_{3,1}y_{3,1} & x_{4,1}y_{4,1} \\[6pt]
  x_{1,2}y_{1,2} & x_{2,2}y_{2,2} & x_{3,2}y_{3,2} & x_{4,2}y_{4,2} \\[6pt]
  x_{1,3}y_{1,3} & x_{2,3}y_{2,3} & x_{3,3}y_{3,3} & x_{4,3}y_{4,3} \\[6pt]
  \vdots & \vdots & \vdots & \vdots \\[6pt]
  x_{1,N}y_{1,N} & x_{2,N}y_{2,N} & x_{3,N}y_{3,N} & x_{4,N}y_{4,N} \\[6pt]
\end{bmatrix},
\end{align}
where $x_{\mu,n} = \pm 1$ and $y_{\mu,n} = \pm 1$ are the values taken by the random variables $X_\mu$ and $Y_\mu$, respectively, in the $n^\text{th}$ repetition of the experiment $E_\mu$, with $\mu=1,2,3,4$, and $n=1,\ldots, N$.

 The sum of the elements of the $\mu^\text{th}$ column gives, by definition, the $N$-sample estimation of the correlation coefficient $\mathbb{E}[X_\mu Y_\mu]$:
\begin{align}\label{b67}
\frac{1}{N} \sum_{n=1}^N  x_{\mu,n}y_{\mu,n} \cong \mathbb{E}[X_\mu Y_\mu],
\end{align}
so that the Bell parameter $S$ is estimated by $S_N$ defined by
\begin{align}\label{b68}
S_N  = \frac{1}{N} \sum_{n=1}^N \bigl( x_{1,n}y_{1,n} + x_{2,n}y_{2,n} + x_{3,n}y_{3,n} - x_{4,n}y_{4,n} \bigr).
\end{align}

According to \eqref{b30},  if a HV theory were valid, the same matrix $V_N$ would be given by,
\begin{widetext}
\begin{align}\label{b70}
V_N^\text{HV} = \frac{1}{N} \begin{bmatrix}
 \vphantom{\Bigl[} {a(\lambda_{1,1})b(\lambda_{1,1})} & a(\lambda_{2,1})b'(\lambda_{2,1}) & {a'(\lambda_{3,1})b(\lambda_{3,1})} & {a'(\lambda_{4,1})b'(\lambda_{4,1})} \\[6pt]
  a(\lambda_{1,2})b(\lambda_{1,2}) & {a(\lambda_{2,2})b'(\lambda_{2,2})} & a'(\lambda_{3,2})b(\lambda_{3,2}) & {a'(\lambda_{4,2})b'(\lambda_{4,2})} \\[6pt]
  {a(\lambda_{1,3})b(\lambda_{1,3})} & a(\lambda_{2,3})b'(\lambda_{2,3}) & a'(\lambda_{3,3})b(\lambda_{3,3}) & a'(\lambda_{4,3})b'(\lambda_{4,3}) \\[6pt]
  \vdots & \vdots & \vdots & \vdots \\[6pt]
  a(\lambda_{1,N})b(\lambda_{1,N}) & {a(\lambda_{2,N})b'(\lambda_{2,N})} & {a'(\lambda_{3,N})b(\lambda_{3,N})} & a'(\lambda_{4,N})b'(\lambda_{4,N}) \\[6pt]
\end{bmatrix}.
\end{align}
\end{widetext}
By definition, the elements of the $N \times 4$ matrix $V_N^\text{HV}$ are equal to either $+1/N$ or $-1/N$, and given by the corresponding elements of $V_N$ in \eqref{b65}. If we sum the $\mu^\text{th}$ column of $V_N^\text{HV}$, we obtain the correlation function \eqref{b30}. Clearly, rearranging the order of the elements in each column will not alter the value of the correlation function. So, we can exploit this freedom to recast $V_N^\text{HV}$ in the new matrix $W_N^\text{HV}$, defined by
\begin{widetext}
\begin{align}\label{b70bis}
W_N^\text{HV} = \frac{1}{N} \begin{bmatrix}
 \vphantom{\Bigl[} 
  {a(\lambda_{1,1})b(\lambda_{1,1})} & a(\lambda_{2,s_1})b'(\lambda_{2,s_1}) & {a'(\lambda_{3,t_1})b(\lambda_{3,t_1})} & {a'(\lambda_{4,f_1})b'(\lambda_{4,f_1})} \\[6pt]
  a(\lambda_{1,2})b(\lambda_{1,2}) & {a(\lambda_{2,s_2})b'(\lambda_{2,s_2})} & a'(\lambda_{3,t_2})b(\lambda_{3,t_2}) & {a'(\lambda_{4,f_2})b'(\lambda_{4,f_2})} \\[6pt]
  {a(\lambda_{1,3})b(\lambda_{1,3})} & a(\lambda_{2,s_3})b'(\lambda_{2,s_3}) & a'(\lambda_{3,t_3})b(\lambda_{3,t_3}) & a'(\lambda_{4,f_3})b'(\lambda_{4,f_3}) \\[6pt]
  \vdots & \vdots & \vdots & \vdots \\[6pt]
  a(\lambda_{1,N})b(\lambda_{1,N}) & {a(\lambda_{2,s_N})b'(\lambda_{2,s_N})} & {a'(\lambda_{3,t_N})b(\lambda_{3,t_N})} & a'(\lambda_{4,f_N})b'(\lambda_{4,f_N}) \\[6pt]
\end{bmatrix}.
\end{align}
\end{widetext}
This matrix has been built in such a way that in the first $M' \leq N$ rows  we have
\begin{align}\label{b75}
\lambda_{1,m} \approx \lambda_{2,s_m} \approx \lambda_{3,t_m} \approx \lambda_{4,f_m},
\end{align}
for all $m = 1, \ldots, M'$. This assumption presupposes that a notion of distance $\| \lambda - \lambda' \|$ between any two points $\lambda$ and $\lambda'$ in $\mathcal{D}$, could be introduced. Clearly, it would be meaningless to be more rigorous about the definition of such a distance since the procedure we are illustrating is purely hypothetical, in the sense that it cannot be materially performed since we are completely ignorant about the parameters $\lambda$ and their distribution $\rho(\lambda)$. What matters is that this procedure \emph{could be performed} if we had a detailed hidden-variable theory. 

Now, let us suppose that in addition to the condition \eqref{b75}, there are  $M \leq M' \leq N$ rows of $W_N^\text{HV}$, such that the random parameters $\lambda$s in Eq.  \eqref{b75}, are close enough to yield
\begin{equation}\label{b190}
\begin{array}{rclll}
&            & \quad a(\lambda_{2,s_m}) & = \; a(\lambda_{m})\;, &    \\[6pt]
& \text{AND} & \quad b(\lambda_{3,t_m}) & = \; b(\lambda_{m}), &  \\[6pt]
&\text{AND}& \quad a'(\lambda_{4,f_m}) & = \; a'(\lambda_{3,t_m}) & = \;   a'(\lambda_{m}) ,  \\[6pt]
&\text{AND}& \quad b'(\lambda_{4,f_m}) & = \; b'(\lambda_{2,s_m}) & = \; b'(\lambda_{m}), 
\end{array}
\end{equation}
where we have redefined $\lambda_m \deff \lambda_{1,m}$.
In this case we would be able to split the sum  in \eqref{b40} in two parts, as follows:
\begin{align}\label{b200}
S_N = \frac{M}{N} \, S_N^{(1)} +  \frac{N-M}{N} \, S_N^{(2)}, 
\end{align}
where 
\begin{align}\label{b210}
S_N^{(1)} & = \frac{1}{M} \sum_{m=1}^M \Bigl[ 
a(\lambda_{m}) b(\lambda_{m}) +
a(\lambda_{m}) b'(\lambda_{m})  \nonumber \\[6pt]
& \phantom{=} \, + a'(\lambda_{m}) b(\lambda_{m}) -
a'(\lambda_{m}) b'(\lambda_{m})  \Bigr],
\end{align}
and
\begin{align}\label{b220}
S_N^{(2)} = & \; \frac{1}{N-M} \sum_{k=1}^{N-M} \Bigl[ 
a(\lambda_{k}) b(\lambda_{k}) +
a(\lambda_{k}') b'(\lambda_{k}')  \nonumber \\[6pt]
&  +
a'(\lambda_{k}'') b(\lambda_{k}'')
- a'(\lambda_{k}''') b'(\lambda_{k}''')  \Bigr].
\end{align}
Clearly, for $S_N^{(1)}$ we have that
\begin{align}\label{b230}
\bigl|S_N^{(1)} \bigr| \leq 2.
\end{align}
However, since $\lambda_k,\lambda_k',\lambda_k'',\lambda_k'''$ are four independent values sampled from the same probability distribution $\rho(\lambda)$, we also have that
\begin{align}\label{b240}
\bigl|S_N^{(2)} \bigr| \leq 4.
\end{align}
Therefore, from (\ref{b200}-\ref{b240}) it follows that
\begin{align}\label{b250}
\bigl| S_N\bigl| \, \leq & \; \frac{M}{N} \, \bigl| S_N^{(1)} \bigr| + \frac{N- M}{N} \, \bigl| S_N^{(2)} \bigr| \nonumber \\[8pt]
 \leq & \;  2 \, \frac{M}{N}  + 4 \, \frac{N- M}{N} = 4 -   2 \frac{M}{N}.
\end{align}
From the law of large numbers \cite{taboga2017}, we expect that (in a probabilistic sense),
\begin{align}\label{b260}
\lim_{N \to \infty} S_N = S.
\end{align}
Therefore, from \eqref{b250} it follows that
\begin{align}\label{b270}
\lim_{N \to \infty} M = N,
\end{align}
so that $4 -  2 \, {M}/{N} \to 2$ in \eqref{b250},  and we recover the theoretical form \eqref{a280} of the CHSH inequality. 
It is interesting to remark that Eq. \eqref{b250} implies that we get $\abs{S_N} > 2$, whenever $M<N$, and we obtain $\abs{S_N} > 2 \sqrt{2}$, whenever $M<(2-\sqrt{2})N \approx 0.6 N$.

Unfortunately, exactly like in the case of the central limit theorem studied in the previous section,  Eq. \eqref{b270} does not give us information about how quickly $M(N)$ converge to $N$ when the latter increases. Of course, this information would be given if we knew a specific HV model. However, this is not the case.

\subsubsection{Discussion}\label{discussion1}

The emission of photon pairs by the source of light in a two-photon correlation experiment naturally occurs a discrete number of times, and the number $N$ of times an experiment can be repeated is necessarily finite. So, if we want a realistic comparison between the predictions of quantum mechanics and those of a hypothetical local realistic hidden-variable theory, we have to take this fact into account. This implies that the actual limit imposed by HV theories on the Bell parameter is not  
\begin{align}\label{b280}
\abs{S} \leq 2,
\end{align}
but rather
\begin{align}\label{b290}
\abs{S_N} \leq 4 - 2 \frac{M}{N},
\end{align}
where the right side of this equation can be any real number between $2$ and $4$, depending on the value of $M = M(N)$. However, there is \emph{no way} to determine the value of the right-hand side of \eqref{b290} from theory alone.

It is important to  emphasize that if we were knowledgeable about hidden variable models, that is, if we knew the nature of the parameters $\lambda$, their distribution $\rho(\lambda)$ and the two functions $a(\alpha, \lambda)$ and $b(\beta, \lambda)$, then the statistical analysis described above would not be purely hypothetical, but it were completely analogous to what is routinely done when using a sample of experimental data to make inferences about a probabilistic model.
The key problem is that we are inevitably ignorant of such models.

\section{Conclusions}\label{conclusions}

In summary, in this paper we have demonstrated two different things. First, no counterfactual reasoning is required to prove the CHSH inequality. Second, the upper bound of this inequality can be greater than $2$, but the numerical value of the actual bound  is not computable because of our unavoidable ignorance about the probability distribution of the hidden variables. To be more specific, the Bell parameter $S$ is bounded by $\abs{S}\leq 2$ for HV theories, where $S$ is the mean value of a linear combination of correlation functions:
\begin{align}\label{b300}
S =  \mathbb{E} \left[ X_1 Y_1  +   X_2 Y_2 + X_3 Y_3 - X_4 Y_4 \right].
\end{align}
However, in a real-world experiment which can be repeated only a finite number $N$ of times, what is measured is not the  mean value $S$, but the sample mean
\begin{align}\label{b310}
S_N & = \frac{1}{N} \sum_{n=1}^N  \Bigl[ X_1(n) Y_1(n)  +   X_2(n) Y_2(n) \nonumber \\
& \phantom{= \frac{1}{N} \sum_{n=1}^N  \Bigl[ } + X_3(n) Y_3(n) - X_4(n) Y_4(n) \Bigr].
\end{align}
The main point of this work is that (unknown) HV theories permit to calculate easily the bounds for the mean $S$, but not for the sample mean $S_N$. Consequently, while we know that  $\abs{S}\leq 2$, we cannot say anything about the actual bounds of $S_N$. We have also shown that both the central limit theorem and the law of large numbers do not help solving this issue because while they guarantee that $S_N \to S$ for $N$ sufficiently large, they do not say anything about the rate of convergence of this limit. So, if one experimentally finds $\abs{S} =  2 \sqrt{2}$, there is a nonzero probability, which, however, is impossible to calculate, that the CHSH inequality is not violated. 

One final note: given the experimental data set collected, for example, in the form of the matrix $V_N$ in \eqref{b65}, a researcher can, of course, perform a full statistical analysis of these data and also infer the rate of convergence of the limit $S_N \to S$. However, the moment he or she wants to compare these data with a given physical model, he or she is faced with an obvious imbalance. In fact, while QM provides a complete model that makes it possible to calculate not only the  mean values of certain measurable quantities but also their probability distributions, generic hidden-variable theories provide, in practice, only  mean values, which are insufficient to infer all the statistical properties of the hidden-variable model. As a matter of fact, the symmetry between quantum and hidden-variable theories of our statistical analysis presented in Sec. \ref{BellPar}, breaks down in the moment the explicit knowledge of the correlation functions $C(\alpha, \beta)$ becomes necessary.  

So, in the end, what the researcher can practically do is just to verify the agreement of the experimental data with the predictions of quantum mechanics, as after all is done in all meaningful quantum physics experiments (except for those checking Bell's theorem!) \cite{Englert2013}.
Because of the relevance of Bell's theorem in practical applications of quantum mechanics, such as quantum computing or quantum cryptography, we believe that these results have the potential to open up new avenues on this important topic.

\appendix

\section{Notation}\label{Anotation}

In this appendix we describe in great detail some notation used in the main text, in order to avoid possible ambiguities. First we introduce the notation suitable for the description of an individual experiment, and then we extend it to the case of four independent experiments.

\subsection{Single experiment}\label{single}

The set of four \emph{independent} experiments $E_\mu(\alpha_\mu, \beta_\mu)$, $(\mu=1, \ldots, 4)$, is described by quantum mechanics as follows. In each individual experiment the  physical system of interest is the pair of photons, called  $A_\mu$ and  $B_\mu$, emitted by the light source. They are characterized by two independent degrees of freedom: the polarization and the path of propagation, all the other degrees of freedom being the same.  
Let us associate the standard bases $\{ \ket{e_+^{(A_\mu)}}, \ket{e_-^{(A_\mu)}}\}$ and $\{ \ket{e_+^{(B_\mu)}}, \ket{e_-^{(B_\mu)}}\}$, to two orthogonal states of linear polarization of photons $A_\mu$ and  $B_\mu$, respectively. By hypothesis, these photons are entangled with respect to the polarization degrees of freedom and each pair can be described by  the state vector $\ket{\psi_\mu} \in \mathcal{H}_\mu= \mathcal{H}^{(A_\mu)} \otimes \mathcal{H}^{(B_\mu)}$, defined by
\begin{align}\label{A10}
\ket{\psi_\mu} & =  \frac{1}{\sqrt{2}} \bigl(  \ket{e_+^{(A_\mu)}} \otimes \ket{e_-^{(B_\mu)}} -  \ket{e_-^{(A_\mu)}} \otimes \ket{e_+^{(B_\mu)}} \bigr) \nonumber \\[6pt]
& =  \frac{1}{\sqrt{2}} \bigl(  \ket{e_+^{(A_\mu)},e_-^{(B_\mu)}} -  \ket{e_-^{(A_\mu)},e_+^{(B_\mu)}} \bigr).
\end{align}

Next, suppose that in the experiment $E_\mu$, Alice and Bob  test the linear polarization state of their photons, using polarizers which can be rotated by the angles $\alpha_\mu$ and $\beta_\mu$ around the direction of propagation of photons $A_\mu$ and  $B_\mu$, respectively.
Each polarizer has two output ports, labeled  $d_{A_\mu +}, d_{A_\mu-}$ for Alice and $d_{B_\mu +}, d_{B_\mu-}$ for Bob. Photons exiting port $d_{A_\mu +}$ ($d_{B_\mu +}$) are linearly polarized along the direction given by the angle $\alpha_\mu$ ($\beta_\mu $) of the rotated polarizer, and photons exiting port $d_{A_\mu -}$ ($d_{B_\mu -}$) are polarized orthogonally to  $\alpha_\mu$ ($\beta_\mu $).

 Let $Z^{(A_\mu)}(\alpha_\mu)$ be an observable that takes the values $+1$ or $-1$, depending on whether the photon exits from port $d_{A_\mu +}$  or port $d_{A_\mu-}$ of Alice's  polarizer. 
Similarly,  $Z^{(B_\mu)}(\beta_\mu)$ is an observable taking the values $+1$ or $-1$, depending on whether the photon exits from port $d_{B_\mu +}$ or port $d_{B_\mu-}$ of Bob's polarizer.  
These two observables can be represented by the Hermitian operators  $\hat{Z}^{(A_\mu)}(\alpha_\mu) \in \mathcal{H}^{(A_\mu)}$ and $\hat{Z}^{(B_\mu)}(\beta_\mu) \in \mathcal{H}^{(B_\mu)}$, which are defined by
\begin{align}\label{A20}
\hat{Z}^{(F)}(\phi)  \deff \proj{\zeta_+^{(F)}(\phi)}{\zeta_+^{(F)}(\phi)} - \proj{\zeta_-^{(F)}(\phi)}{\zeta_-^{(F)}(\phi)},  
\end{align}
where
\begin{equation}\label{A30}
\begin{split}
\ket{\zeta_+^{(F)}(\phi)} & = \cos \phi \ket{e_+^{(F)}} + \sin \phi \ket{e_-^{(F)}}, \\[6pt]
\ket{\zeta_-^{(F)}(\phi)} & = -\sin \phi \ket{e_+^{(F)}} + \cos \phi \ket{e_-^{(F)}},
\end{split}
\end{equation}
with $\ket{\zeta_\pm^{(F)}(0)}= \ket{e_\pm^{(F)}} $, and $(F,\phi) = (A_\mu, \alpha_\mu)$, or $(F,\phi) = (B_\mu, \beta_\mu)$. 
By definition,
\begin{align}\label{A40}
\hat{Z}^{(F)}(\phi)  \ket{\zeta_\pm^{(F)}(\phi)} = \pm \ket{\zeta_\pm^{(F)}(\phi)} .
\end{align}

The two-photon Hilbert space $\mathcal{H}_\mu$ is spanned by the basis vectors $\ket{k_\mu, l_\mu} $, defined by
\begin{align}\label{A90}
\ket{k_\mu, l_\mu} \deff \ket{\zeta_{k_\mu}^{(A_\mu)}(\alpha_\mu)} \otimes \ket{\zeta_{l_\mu}^{(B_\mu)}(\beta_\mu)}.
\end{align}
By definition, these vectors form an orthonormal and complete basis in the space $\mathcal{H}_\mu$, that is
\begin{equation}\label{A100}
\begin{split}
\brak{k_\mu, l_\mu}{k_\nu, l_\nu} & = \delta(k_\mu ,k_\nu)\delta(l_\mu ,l_\nu), \\[6pt]
\sum_{k_\mu, l_\mu = \pm 1} \proj{k_\mu, l_\mu}{k_\mu, l_\mu} & = \hat{I}_\mu, \quad \; (\mu, \nu = 1, \ldots, 4),
\end{split}
\end{equation}
where $\hat{I}_\mu \in \mathcal{H}_\mu$ is the identity operator, and  $ \delta(i,j)$ denotes the Kronecker delta function defined by
\begin{align}\label{A110}
\delta(i,j) = \begin{cases}
0 &\text{if } i \neq j,   \\
1 &\text{if } i=j.   \end{cases}
\end{align}

It is convenient to extend $\hat{Z}^{(A_\mu)}(\alpha_\mu)$ and $\hat{Z}^{(B_\mu)}(\beta_\mu)$ to the Hilbert space $\mathcal{H}_\mu$, by defining
\begin{subequations}\label{A50}
\begin{align}
\hat{X}_\mu(\alpha_\mu) & \deff \hat{Z}^{(A_\mu)}(\alpha_\mu) \otimes \hat{I}^{(B_\mu)} , \label{A50A} \\[6pt]
\hat{Y}_\mu(\beta_\mu) & \deff \hat{I}^{(A_\mu)} \otimes \hat{Z}^{(B_\mu)}(\beta_\mu) . \label{A50B}
\end{align}
\end{subequations}
From \eqref{A40} and \eqref{A90}, it follows that
\begin{subequations}\label{A55}
\begin{align}
\hat{X}_\mu(\alpha_\mu) \ket{k_\mu, l_\mu} & =  k_\mu \ket{k_\mu, l_\mu} , \label{A55A} \\[6pt]
\hat{Y}_\mu(\beta_\mu) \ket{k_\mu, l_\mu} & =  l_\mu \ket{k_\mu, l_\mu} , \label{A55B}
\end{align}
\end{subequations}
where $k_\mu, l_\mu = \pm 1$. Moreover, 
\begin{align}\label{A60}
\bigl[ \hat{X}_\mu(\alpha_\mu) ,  \hat{Y}_\mu(\beta_\mu)\bigr] = 0,
\end{align}
for all angles $\alpha_\mu$ and $\beta_\mu$. From a physical point of view, this simple relation has a very important meaning: the measurements performed on photons $A_\mu$ and $B_\mu$ are independent of each other. In other words, any polarization test made by Alice has not influence on  any other polarization measurement made by Bob, and vice versa,  regardless of the state of the two photons. There is no room for further discussion on this point, as also made clear in Sec. 7 of Ref. \cite{Englert2013}.

Consider now the result of a measurement of the correlation between the values assumed by the observables $\hat{Z}^{(A_\mu)}(\alpha_\mu)$ and $\hat{Z}^{(B_\mu)}(\beta_\mu)$ in a single run of the experiment $E_\mu$. Such correlation is quantified by the expectation value of the correlation operator 
\begin{align}\label{A70}
\hat{C}_\mu(\alpha_\mu, \beta_\mu) \deff \hat{X}_\mu(\alpha_\mu)\hat{Y}_\mu(\beta_\mu) \in \mathcal{H}_\mu,
\end{align}
with respect the state $\ket{\psi_\mu}$. By construction,
\begin{align}\label{A80}
\hat{C}_\mu(\alpha_\mu, \beta_\mu)  \ket{k_\mu, l_\mu} = k_\mu l_\mu \ket{k_\mu, l_\mu},
\end{align}
with $k_\mu, l_\mu = \pm 1$.

\subsection{Four experiments}\label{four}

Consider now the four \emph{independent} experiments $E_1, E_2, E_3$ and $E_4$ together. Each of the four sources emits a pair of photons in the state $\ket{\psi_\mu}$, so that the overall four-pair state is given by \cite{ADENIER2001},
\begin{align}\label{A120}
\ket{\Psi} = \ket{\psi_1} \otimes \ket{\psi_2} \otimes \ket{\psi_3} \otimes \ket{\psi_4} \in \mathcal{H},
\end{align}
where
\begin{align}\label{A130}
\mathcal{H} = \mathcal{H}_1 \otimes \mathcal{H}_2 \otimes \mathcal{H}_3 \otimes \mathcal{H}_4.
\end{align}
The Hilbert space $\mathcal{H}$ is spanned by  the 256 basis vectors
\begin{align}\label{A150}
\Bigl\{ \bigotimes_{\mu=1}^4 \ket{k_\mu, l_\mu} \Bigr\} & = \Bigl\{ \ket{k_1, l_1}\ket{k_2, l_2}\ket{k_3, l_3}\ket{k_4, l_4} \Bigr\} \nonumber \\[6pt]
& \deff \Bigl\{ \ket{K, L} \Bigr\}.
\end{align}
with $k_1, l_1, \ldots, k_4,l_4 = \pm 1$. From \eqref{A100} it follows that  these vectors form an orthonormal and complete basis in the space $\mathcal{H}$, that is
\begin{equation}\label{A100bis}
\begin{split}
\brak{K,L}{K',L'} & = \delta(K,K')\delta(L,L'), \\[6pt]
\sum_{K,L} \proj{K,L}{K,L} & = \hat{I}, 
\end{split}
\end{equation}
where $\hat{I} \in \mathcal{H}$ is the identity operator,
\begin{align}\label{A152}
\hat{I}  & = \bigotimes_{\mu=1}^4 \hat{I}_\mu \nonumber \\[6pt]
& = \bigotimes_{\mu=1}^4 \Bigl\{ \sum_{k_\mu, l_\mu = \pm 1} \proj{k_\mu, l_\mu}{k_\mu, l_\mu} \Bigr\} ,
\end{align}
and $\delta(K,K')\delta(L,L')$ is a shorthand for
\begin{align}\label{A154}
\delta(K,K')\delta(L,L') = \prod_{\mu=1}^4 \delta(k_\mu, k_\mu')\delta(l_\mu, l_\mu').
\end{align}

Next, using \eqref{A70} we define the ``Bell operator'' $\hat{T} \in \mathcal{H}$, as
\begin{align}\label{A140}
\hat{T} & \deff \hat{C}_1(\alpha, \beta) \otimes \hat{I}_2 \otimes \hat{I}_3 \otimes \hat{I}_4 \nonumber \\[6pt]
& \phantom{\deff} +  \hat{I}_1 \otimes \hat{C}_2(\alpha, \beta') \otimes \hat{I}_3 \otimes \hat{I}_4 \nonumber \\[6pt]
& \phantom{\deff} + \hat{I}_1 \otimes \hat{I}_2 \otimes \hat{C}_3(\alpha', \beta) \otimes \hat{I}_4 \nonumber \\[6pt]
& \phantom{\deff} + \hat{I}_1 \otimes \hat{I}_2 \otimes \hat{I}_3 \otimes \hat{C}_4(\alpha', \beta') \nonumber \\[6pt]
& \deff \hat{T}_1 + \hat{T}_2 + \hat{T}_3 - \hat{T}_4.
\end{align}
From (\ref{A80},\ref{A150}) and \eqref{A140}, it follows that 
\begin{align}\label{A160}
\hat{T}_\nu \ket{K, L} = k_\nu l_\nu \ket{K, L}, 
\end{align}
with $\nu =1,\ldots,4$, which implies
\begin{align}\label{A170}
\hat{T} \ket{K, L} = \left( k_1 l_1 + k_2 l_2 + k_3 l_3 - k_4 l_4 \right) \ket{K, L}. 
\end{align}
Since
\begin{align}\label{A180}
k_1 l_1 + k_2 l_2 + k_3 l_3 - k_4 l_4 = 0, \pm 2, \pm 4,
\end{align}
when $k_1, l_1, \ldots, k_4,l_4 = \pm 1$, then  $\hat{T}$ possesses only $5$ distinct eigenvalues, multiply degenerate.  Specifically, the degree of degeneracy $g$ is given by, 
\begin{align}\label{A190}
g[\pm 4] = 16, \qquad g[\pm 2] = 64, \qquad g[0] = 96.
\end{align}
The detailed  random variable $T$ representation of the operator $\hat{T}$ with respect to the vector state $\ket{\Psi}$, will be given in Appendix \ref{opT}, here we report only the main formulas. The expectation value of $T$ is given by,
\begin{align}\label{A200}
\mathbb{E}[T]  = & -\cos[2(\alpha -\beta)] - \cos[2(\alpha -\beta')] \nonumber \\[6pt]
 &  - \cos[2(\alpha' -\beta)] +  \cos[2(\alpha' -\beta')],
\end{align}
as it should be. The variance of $T$ is
\begin{align}\label{A210}
\mathbb{V}[T] &  = \mathbb{E}\bigl[\left( T -\mathbb{E}[T] \right)^2 \bigr]  \nonumber \\[6pt]
&  = 2 - \frac{1}{2} \Bigl\{ \cos [4 (\alpha -\beta )] + \cos [4 (\alpha -\beta' )] \nonumber \\[6pt]
& \phantom{=  .}  + \cos [4 (\alpha' -\beta )]  + \cos [4 (\alpha' -\beta' )]   \Bigr\},
\end{align}
in agreement with \eqref{probT40}.

\section{QM  probability distributions}\label{prob}

In this Appendix we calculate explicitly the probability distribution $p_{kl}^\text{QM}(\alpha_\mu,\beta_\mu) $ that we have introduced in Sec. \ref{deriveCHSH}. However, first we provide for a short recap of some formulas from random variables theory, which we will need later.

\subsection{Short recap of some random variables theory formulas}\label{briefprob}

We briefly recall that from von Neumann's spectral theorem  \cite{vonNeumann2018,aiello_arXiv.2110.12930}, it follows that  given a self-adjoint operator $\hat{X}$ and a normalized state vector $\ket{\psi}$, there is a unique random variable $X$ associated with $\hat{X}$ and $\ket{\psi}$, which is distributed according to the probability density function $f^{\ket{\psi}}_X(x)$ defined by,
\begin{align}\label{B10}
f^{\ket{\psi}}_X(x) = \mean{\psi}{\delta \bigl( x \hat{I} - \hat{X} \bigr)}{\psi},
\end{align}
where $x \in \mathbb{R}$ is one of the values assumed by $X$ when an experiment is performed (see, e.g., sec. 3-1-2 in \cite{ItzZub}, and problem 4.3 in \cite{Coleman2019}). When $\hat{X}$ possesses a discrete spectrum, that is $\hat{X}_n \ket{x_n} = x_n \ket{x_n}$, with $n \in \mathbb{N}$, then \eqref{b10} can be rewritten as
\begin{align}\label{B20}
f^{\ket{\psi}}_X(x) & = \sum_{n \in \mathbb{N}} \delta(x - x_n) \abs{\brak{x_n}{\psi}}^2 \nonumber \\[6pt]
& \deff \sum_{n \in \mathbb{N}} \delta(x - x_n)p_n,
\end{align}
where $p_n=  \abs{\brak{x_n}{\psi}}^2$, is the so-called probability mass function of the random variable $X$.

Thus, we write $X \sim f^{\ket{\psi}}_X(x)$, and we can  calculate the expectation value of any regular function $F(\hat{X}) $ of $\hat{X}$ with respect to $| \psi \rangle$, either as $\langle F(\hat{X}) \rangle_\psi  = \langle \psi | F(\hat{X}) | \psi \rangle$, or as $\langle F(\hat{X}) \rangle_\psi  = \mathbb{E}[F(X)]$, where $\mathbb{E}[F(X)]$ denotes the expected value  of the random variable $F(X)$, calculated as
\begin{align}\label{B30}
\mathbb{E}[F(X)]  & = \int_{\mathbb{R}}  F(x) \, f^{\ket{\psi}}_X(x) \, \mathrm{d} x \nonumber \\[6pt]
& = \sum_{n\in \mathbb{N}} F(x_n)\, p_n,
\end{align}
where the equation in the second line applies when $\hat{X}$ has a discrete spectrum with eigenvalues $x_n$.

\subsection{Calculation of $p_{k_\mu l_\mu}^\mathrm{QM}(\alpha_\mu,\beta_\mu) $}\label{probQM}

From \eqref{A50} and \eqref{B10}, it follows that given the operators $\hat{X}_\mu(\alpha_\mu)$ and $\hat{Y}_\mu(\beta_\mu)$, and given the vector state $\ket{\psi_\mu}$, we can define the two random variables $X_\mu$ and $Y_\mu$, distributed according to
\begin{align}\label{B40}
X_\mu \sim f^{\ket{\psi_\mu}}_{X_\mu}(x) , \qquad \text{and} \qquad Y_\mu \sim f^{\ket{\psi_\mu}}_{Y_\mu}(y),
\end{align}
where
\begin{subequations}\label{B50}
\begin{align}
f^{\ket{\psi_\mu}}_{X_\mu}(x) & = \mean{\psi_\mu}{\delta \bigl( x \hat{I}_\mu - \hat{X}_\mu(\alpha_\mu) \bigr)}{\psi_\mu} , \label{B50A} \\[6pt]
f^{\ket{\psi_\mu}}_{Y_\mu}(y) & = \mean{\psi_\mu}{\delta \bigl( y \hat{I}_\mu - \hat{Y}_\mu(\beta_\mu) \bigr)}{\psi_\mu}  . \label{B50B}
\end{align}
\end{subequations}
For illustration purposes, we will perform an explicit calculation only for $f^{\ket{\psi_\mu}}_{X_\mu}(x)$. Inserting the identity operator \eqref{A100} into \eqref{B50A}, we obtain
\begin{widetext}
\begin{align}\label{B60}
f^{\ket{\psi_\mu}}_{X_\mu}(x) & = \bra{\psi_\mu}{\delta \bigl( x \hat{I}_\mu - \hat{X}_\mu(\alpha_\mu) \bigr)}\left( \sum_{k_\mu, l_\mu = \pm 1} \proj{k_\mu, l_\mu}{k_\mu, l_\mu} \right)\ket{\psi_\mu} \nonumber \\[6pt]
 & = \sum_{k_\mu = \pm 1} \delta \bigl( x  - k_\mu \bigr) \sum_{ l_\mu = \pm 1}\abs{\brak{k_\mu, l_\mu}{\psi_\mu}}^2 \nonumber \\[6pt]
 & \deff  \sum_{k_\mu = \pm 1} \delta \bigl( x  - k_\mu \bigr) p_{k_\mu}^\text{QM}, 
\end{align}
\end{widetext}
where \eqref{A55A} has been used, and we have defined 
\begin{align}\label{B70}
p_{k_\mu}^\text{QM} & \deff  \sum_{ l_\mu = \pm 1}\abs{\brak{k_\mu, l_\mu}{\psi_\mu}}^2 \nonumber \\[6pt]
 & = \bra{\psi_\mu} \left( \proj{k_\mu}{k_\mu} \otimes \sum_{ l_\mu = \pm 1} \proj{l_\mu}{l_\mu} \right) \ket{\psi_\mu} \nonumber \\[6pt]
 & = \bra{\psi_\mu} \left( \proj{k_\mu}{k_\mu} \otimes \hat{I}^{(B_\mu)} \right) \ket{\psi_\mu} .
\end{align}
Substituting \eqref{A10} into \eqref{B70}, we obtain
\begin{align}\label{B80}
p_{k_\mu}^\text{QM} &  = \frac{1}{2} \left[ \abs{\brak{e_+^{(A_\mu)}}{k_\mu}}^2 + \abs{\brak{e_-^{(A_\mu)}}{k_\mu}}^2 \right] \nonumber \\[6pt]
 & = \frac{1}{2}\bra{k_\mu} \underbrace{\biggl[ \proj{e_+^{(A_\mu)}}{e_+^{(A_\mu)}} + \proj{e_-^{(A_\mu)}}{e_-^{(A_\mu)}} \biggr]}_{= \; \hat{I}_\mu} \ket{k_\mu}\nonumber \\
 & = \frac{1}{2}\brak{k_\mu}{k_\mu} \nonumber \\[6pt]
 & = \frac{1}{2}.
\end{align}

Since the two operators $\hat{X}_\mu(\alpha_\mu)$ and $\hat{Y}_\mu(\beta_\mu)$ commute, it is possible to calculate the joint probability density function $f^{\ket{\psi_\mu}}_{X_\mu Y_\mu}(x,y)$, defined by 
\begin{widetext}
\begin{align}\label{B90}
f^{\ket{\psi_\mu}}_{X_\mu Y_\mu}(x,y) &  = \mean{\psi_\mu}{\delta \bigl( x \hat{I}_\mu - \hat{X}_\mu(\alpha_\mu) \bigr)\delta \bigl( y \hat{I}_\mu - \hat{Y}_\mu(\beta_\mu) \bigr)}{\psi_\mu} \nonumber \\[6pt]
 & =  \bra{\psi_\mu}{\delta \bigl( x \hat{I}_\mu - \hat{X}_\mu(\alpha_\mu) \bigr) \delta \bigl( y \hat{I}_\mu - \hat{Y}_\mu(\beta_\mu) \bigr)}\left( \sum_{k_\mu, l_\mu = \pm 1} \proj{k_\mu, l_\mu}{k_\mu, l_\mu} \right)\ket{\psi_\mu} \nonumber \\[6pt]
 & =  \sum_{k_\mu, l_\mu = \pm 1} \delta \bigl( x  - k_\mu \bigr)\delta \bigl( y  - l_\mu \bigr) \abs{\brak{k_\mu, l_\mu}{\psi_\mu}}^2
 \nonumber \\[6pt]
 & \deff \sum_{k_\mu, l_\mu = \pm 1} \delta \bigl( x  - k_\mu \bigr)\delta \bigl( y  - l_\mu \bigr)  p_{k_\mu l_\mu}^\text{QM}(\alpha_\mu, \beta_\mu),
\end{align}
\end{widetext}
where we have defined 
\begin{align}\label{B100}
p_{k_\mu l_\mu}^\text{QM}(\alpha_\mu, \beta_\mu) \deff \abs{\brak{k_\mu, l_\mu}{\psi_\mu}}^2, \quad (k_\mu, l_\mu = \pm 1).
\end{align}
Substituting \eqref{A10} and \eqref{A90} into \eqref{B100}, we find
\begin{align}\label{B110}
p_{k_\mu l_\mu}^\text{QM}(\alpha_\mu, \beta_\mu) = \frac{1}{4} \left\{ 1 - k_\mu l_\mu\cos \bigl[ 2 \left( \alpha_\mu - \beta_\mu \right)\bigr]\right\}.
\end{align}
Using this equation, it is easy to obtain again the marginal distribution \eqref{B80} as,
\begin{align}\label{B120}
p_{k_\mu}^\text{QM} = \sum_{l_\mu = \pm 1} p_{k_\mu l_\mu}^\text{QM}(\alpha_\mu, \beta_\mu) = \frac{1}{2} .
\end{align}

\section{Probabilities $p_n$ in Eq. \eqref{probT20}}\label{pn}

In this Appendix we adopt the shorthand $C_1 = C(\alpha, \beta)$, $C_2 = C(\alpha, \beta')$ , $C_3 = C(\alpha', \beta)$, and $C_4 = C(\alpha', \beta')$. A straightforward calculation gives:
\begin{widetext}
\begin{equation}\label{PN10}
\begin{split}
P(T = 0) & = \frac{3}{8} - \frac{1}{8} \left( C_1 C_2 + C_1 C_3 - C_1 C_4 + C_2 C_3 - C_2 C_4 - C_3 C_4 \right) -\frac{3}{8} C_1 C_2 C_3 C_4 , \\[6pt]
P(T = \pm 2) & = \frac{1}{4} \pm \frac{1}{8} \left( C_1 + C_2 + C_3 - C_4 \right) \mp \left( C_1 C_2 C_3 - C_1 C_2 C_4 - C_1 C_3 C_4 - C_2 C_3 C_4 \right) +\frac{1}{4} C_1 C_2 C_3 C_4,  \\[6pt]
P(T = \pm 4) & = \frac{1}{16} \pm \frac{1}{16}  \left( C_1 + C_2 + C_3 - C_4 \right)  + \frac{1}{16} \left(  C_1 C_2 + C_1 C_3 - C_1 C_4 + C_2 C_3 - C_2 C_4 - C_3 C_4 \right) \\[6pt]
 & \phantom{=.}  \pm \frac{1}{16} \left( C_1 C_2 C_3 - C_1 C_2 C_4 - C_1 C_3 C_4 - C_2 C_3 C_4 \right) - \frac{1}{16} C_1 C_2 C_3 C_4 .
\end{split}
\end{equation}
\end{widetext}

\section{Random variable representation of the quantum Bell operator $\hat{T}$}\label{opT}

In this section we calculate and study the random variable $T$ associated with the operator $\hat{T}$ and the vector state $\ket{\Psi}$, defined by Eqs. \eqref{A140} and \eqref{A120}, respectively.  Using \eqref{B10} we can calculate
\begin{align}\label{E10}
f^{\ket{\Psi}}_T(t) & = \mean{\Psi}{\delta \bigl( t \hat{I} - \hat{T} \bigr)}{\Psi} \nonumber \\[6pt]
 & = \sum_{K,L} \bra{\Psi}\delta \bigl( t \hat{I} - \hat{T} \bigr)\ket{K,L} \brak{K,L}{\Psi},
\end{align}
where \eqref{A100bis} has been used. From \eqref{A170} it follows that
\begin{multline}\label{E20}
\delta \bigl( t \hat{I} - \hat{T} \bigr)\ket{K,L} \\[6pt] = \delta \bigl( t -k_1 l_1 - k_2 l_2 - k_3 l_3 + k_4 l_4  \bigr)\ket{K,L} .
\end{multline}
Substituting \eqref{E20} into \eqref{E10}, we obtain
\begin{multline}\label{E30}
f^{\ket{\Psi}}_T(t) \\[6pt] =  \sum_{K,L}  \delta \bigl( t -k_1 l_1 - k_2 l_2 - k_3 l_3 + k_4 l_4  \bigr) \abs{\brak{K,L}{\Psi}}^2,
\end{multline}
and we use \eqref{A120} and \eqref{A150} to calculate,
\begin{align}\label{E40}
\abs{\brak{K,L}{\Psi}}^2 & = \prod_{\mu=1}^4 \abs{\brak{k_\mu, l_\mu}{\psi_\mu}}^2 \nonumber \\[6pt]
& = \prod_{\mu=1}^4  p_{k_\mu l_\mu}^\text{QM}(\alpha_\mu, \beta_\mu) ,
\end{align}
where \eqref{B100} has been used. So, being the four experiments independent, the probability distribution of $T$ is factorable, as expected. This implies that
\begin{widetext}
\begin{align}\label{E50}
\mathbb{E}[T] & = \int_\mathbb{R}  t \, f^{\ket{\Psi}}_T(t) \, \di t \nonumber \\[6pt]
& = \sum_{k_1,l_1 = \pm 1} \sum_{k_2,l_2 = \pm 1} \sum_{k_3,l_3 = \pm 1} \sum_{k_4,l_4 = \pm 1} \prod_{\mu=1}^4  p_{k_\mu l_\mu}^\text{QM}(\alpha_\mu, \beta_\mu)
 \int_\mathbb{R}  t \, \delta \bigl( t -k_1 l_1 - k_2 l_2 - k_3 l_3 + k_4 l_4  \bigr) \, \di t \nonumber \\[6pt]
& = \sum_{k_1,l_1 = \pm 1} \sum_{k_2,l_2 = \pm 1} \sum_{k_3,l_3 = \pm 1} \sum_{k_4,l_4 = \pm 1}  p_{k_1 l_1}^\text{QM}(\alpha, \beta)
p_{k_2 l_2}^\text{QM}(\alpha, \beta') p_{k_3 l_3}^\text{QM}(\alpha', \beta)p_{k_4 l_4}^\text{QM}(\alpha', \beta')
 \bigl(k_1 l_1 + k_2 l_2 + k_3 l_3 - k_4 l_4  \bigr)  \nonumber \\[6pt]
& = \sum_{\mu=1}^4 s_\mu  \left[ \sum_{k_\mu, l_\mu = \pm 1} k_\mu l_\mu \, p_{k_\mu l_\mu}^\text{QM}(\alpha_\mu, \beta_\mu) \right], 
\end{align}
where we have introduced the useful parameters $s_1=s_2=s_3 = 1$, and $s_4 = -1$.
\end{widetext}
Substituting \eqref{a170} into \eqref{E50}, we recover the well known result
\begin{align}\label{E60}
\mathbb{E}[T] & = \sum_{\mu=1}^4  s_\mu  C^\text{QM}(\alpha_\mu, \beta_\mu) \nonumber \\[6pt]
& = -\cos[2(\alpha -\beta)] - \cos[2(\alpha -\beta')] \nonumber \\[6pt]
& \phantom{=.}  - \cos[2(\alpha' -\beta)] +  \cos[2(\alpha' -\beta')].
\end{align}

An important parameter characterizing the probability distribution $f_X(x)$ of a random variable $X$, is the variance $\mathbb{V}[X]$, defined by
\begin{align}\label{E70}
\mathbb{V}[X]   = \mathbb{E}[X^2] - (\mathbb{E}[X])^2.
\end{align}
In our case a straightforward calculation gives
\begin{align}\label{E80}
\mathbb{V}[T] &  = 2 - \frac{1}{2} \Bigl\{ \cos [4 (\alpha -\beta )] + \cos [4 (\alpha -\beta' )] \nonumber \\[6pt]
& \phantom{=  2 - \frac{1}{2} \Bigl\{}  + \cos [4 (\alpha' -\beta )]  + \cos [4 (\alpha' -\beta' )]   \Bigr\}.
\end{align}

\begin{figure}[h!]
  \centering
  \includegraphics[scale=3,clip=false,width=0.9\columnwidth,trim = 0 0 0 0]{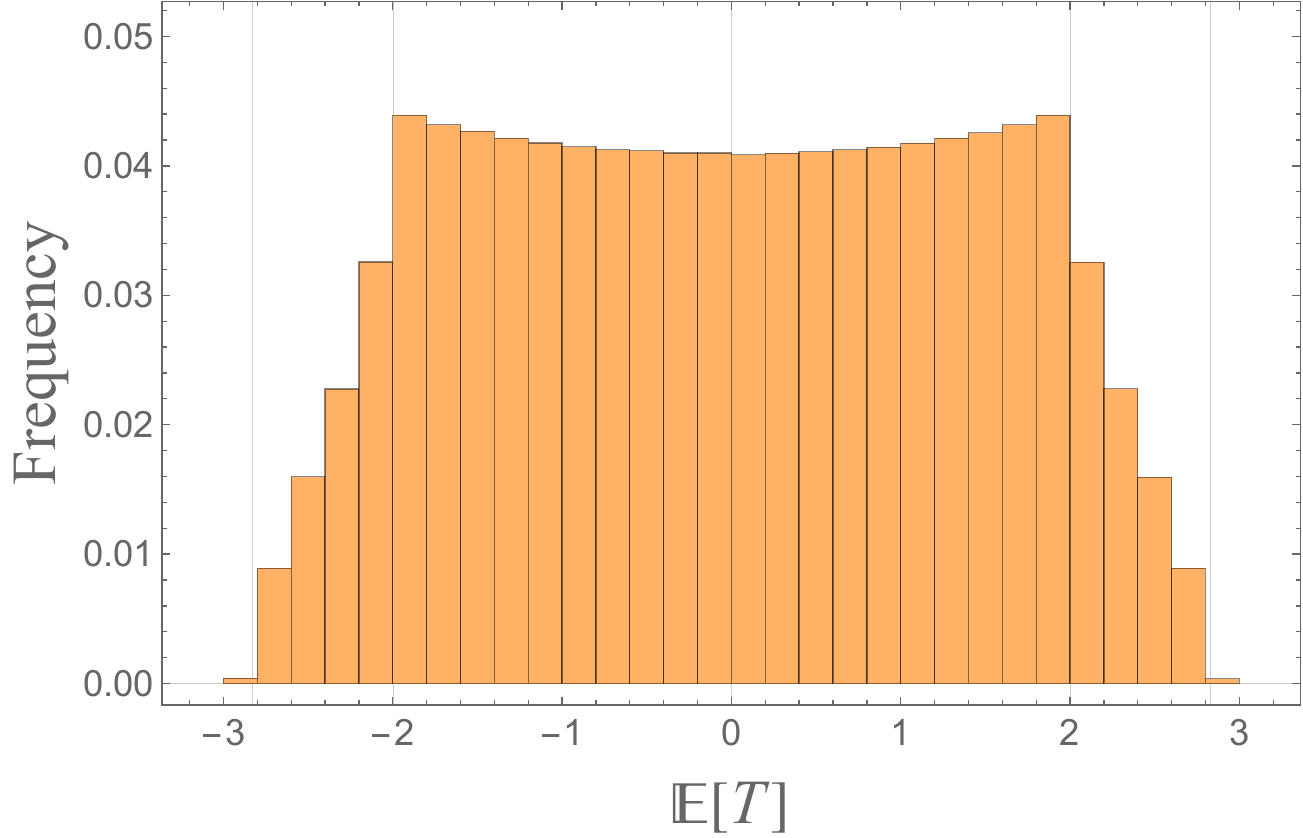}
  \caption{Frequency of occurrence of the values of the mean $\mathbb{E}[T]$, which are reported on the horizontal axis, when the four angles $\alpha, \beta, \alpha', \beta'$, are uniformly  sampled at random in the interval $[0,\pi)$. The histogram was constructed from a sample of $2^{24}$ random numbers. The vertical lines mark the values $\pm 2 \sqrt{2}$, $\pm 2$, and $0$.}\label{figE1}
\end{figure}

\begin{figure}[h!]
  \centering
  \includegraphics[scale=3,clip=false,width=0.9\columnwidth,trim = 0 0 0 0]{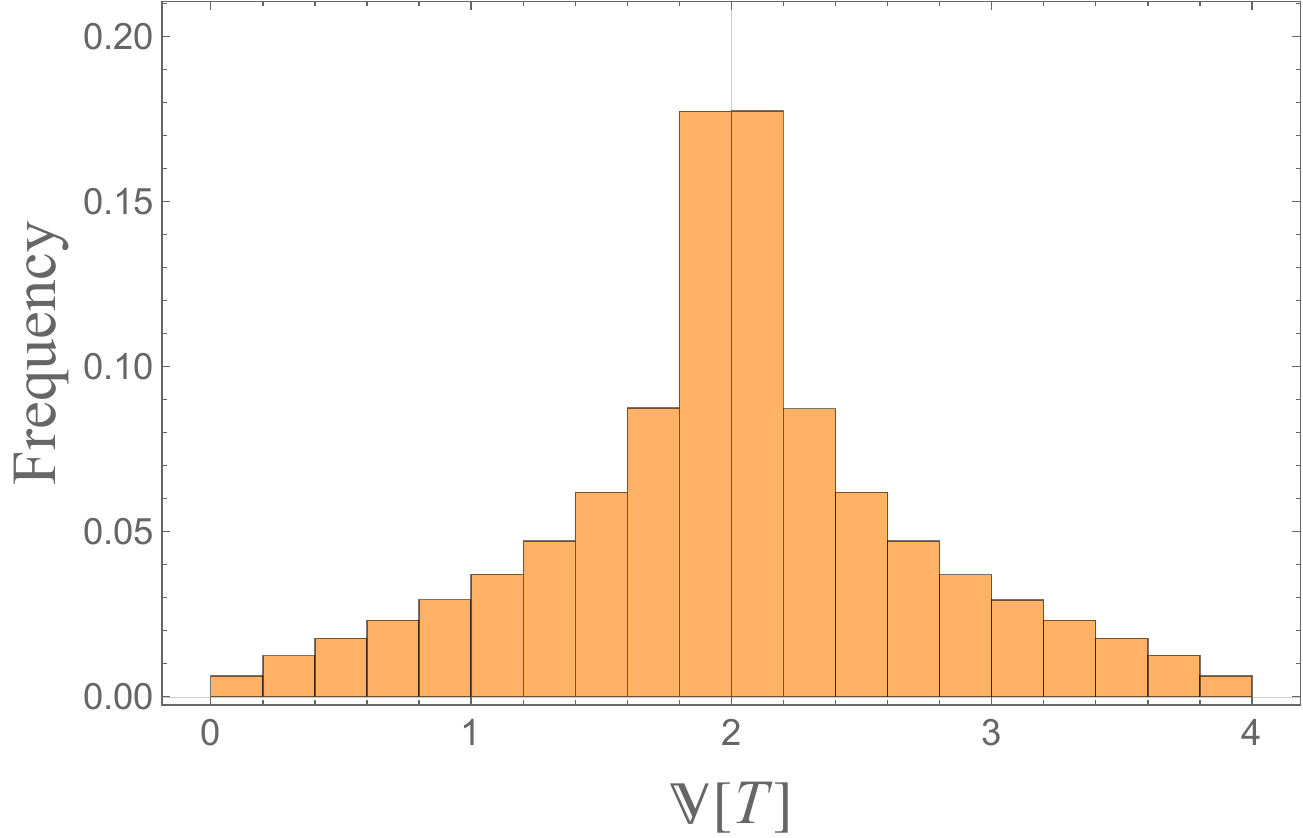}
  \caption{Frequency of occurrence of the values of the variance $\mathbb{V}[T]$, which are reported on the horizontal axis, when the four angles $\alpha, \beta, \alpha', \beta'$, are uniformly  sampled at random in the interval $[0,\pi)$. The histogram was constructed from a sample of $2^{24}$ random numbers.}\label{figE2}
\end{figure}

It is instructive to visualize the range of $\mathbb{E}[T]$ and $\mathbb{V}[T]$ by randomly sampling the four angles $\alpha, \beta, \alpha', \beta'$, uniformly in the interval $[0,\pi)$, and plotting the frequency of occurrence of the values of $\mathbb{E}[T]$ and $\mathbb{V}[T]$. This is shown in Figs. \ref{figE1} and \ref{figE2}, respectively.
From Fig. \ref{figE1} it is interesting to note how the values of $\mathbb{E}[T]$ in the interval $[-2,2]$, are almost uniformly distributed, only slightly peaked at $\mathbb{E}[T] = \pm 2$. Outside this interval the frequency of occurrence of $\mathbb{E}[T]$ drops significatively, being actually quite low for $\mathbb{E}[T] = \pm 2 \sqrt{2}$. This means that values of $\abs{\mathbb{E}[T]}$ bigger than $2$, occur for a small range of Alice and bob's polarizers orientations.  
The plot of the variance $\mathbb{V}[T]$ in Fig. \ref{figE2}, shows that $\mathbb{V}[T] = 2$ is the most likely value. The frequency of occurrence of $\mathbb{V}[T] \neq 2$, drops quickly to zero when we move away from  $\mathbb{V}[T] = 2$.

\subsection{Four-dimensional Hilbert space representation of the operator $\hat{T}$}\label{fourD}

The Bell operator $\hat{T}$ in \eqref{A140}, is defined in the 256-dimensional Hilbert space $\mathcal{H} = \mathcal{H}_1 \otimes \mathcal{H}_2 \otimes \mathcal{H}_3 \otimes \mathcal{H}_4$, because quantum mechanics forbids the \emph{separate} and \emph{simultaneous} measurement of the four correlation functions in the expression  \eqref{a140} of $S(\alpha, \beta, \alpha', \beta')$. However,  it is not really necessary to measure such functions separately \emph{and} simultaneously. This kind of situation occurs quite frequently in quantum mechanics. For example, one can measure the discrete energy spectrum of a harmonic oscillator without measuring separately and simultaneously both the kinetic and potential energy, which would be impossible to do because position and momentum operators do not commute.
In principle, it would not be impossible to design an experiment whose outcomes are the  eigenvalues of the \emph{whole} operator $\hat{S} \in \mathcal{H}^{(A)} \otimes \mathcal{H}^{(B)}$, defined by
\begin{align}\label{d280}
\hat{S} & \deff \hat{Z}^{(A)}(\alpha) \otimes \hat{Z}^{(B)}(\beta) + \hat{Z}^{(A)}(\alpha) \otimes \hat{Z}^{(B)}(\beta') \nonumber \\[6pt]
& \phantom{=}  + \hat{Z}^{(A)}(\alpha') \otimes \hat{Z}^{(B)}(\beta) - \hat{Z}^{(A)}(\alpha') \otimes \hat{Z}^{(B)}(\beta').
\end{align}
From \eqref{d280} it follows that $\hat{S}$ is defined in the 4-dimensional two-photon Hilbert space.  In this equation
\begin{align}\label{A20bis}
\hat{Z}^{(F)}(\phi)  \deff \proj{\zeta_+^{(F)}(\phi)}{\zeta_+^{(F)}(\phi)} - \proj{\zeta_-^{(F)}(\phi)}{\zeta_-^{(F)}(\phi)},  
\end{align}
where
\begin{equation}\label{A30bis}
\begin{split}
\ket{\zeta_+^{(F)}(\phi)} & = \cos \phi \ket{e_+^{(F)}} + \sin \phi \ket{e_-^{(F)}}, \\[6pt]
\ket{\zeta_-^{(F)}(\phi)} & = -\sin \phi \ket{e_+^{(F)}} + \cos \phi \ket{e_-^{(F)}}, \\[6pt]
\end{split}
\end{equation}
with $\ket{\zeta_\pm^{(F)}(0)}= \ket{e_\pm^{(F)}} $, and $(F,\phi) = (A, \alpha)$, or $(F,\phi) = (B_, \beta)$. 

The operator $\hat{S}$ is self-adjoint by construction,  therefore it could legitimately represent an actual observable.  Using  von Neumann's spectral theorem  \cite{vonNeumann2018,aiello_arXiv.2110.12930}, it is not difficult to calculate the probability density function $f_S^{\ket{\Psi}}(s)$ for the random variable $S$ associated with the operator $\hat{S}$ and the state vector $\ket{\Psi}$. 
As usual, this density is defined by
\begin{align}\label{d290}
f_S^{\ket{\Psi}}(s) = \mean{\Psi}{\delta \bigl( s \, \hat{I}^{(A)} \otimes \hat{I}^{(B)} - \hat{S} \bigr)}{\Psi}.
\end{align}
A straightforward calculation gives
\begin{align}\label{d300}
f_S^{\ket{\Psi}}(s) & = \frac{1}{2} \left(1 + \frac{E}{s_0} \right)\delta(s-s_0) \nonumber \\[6pt]
& \phantom{=} + \frac{1}{2} \left(1 - \frac{E}{s_0} \right)\delta(s+s_0),
\end{align}
where we have defined
\begin{align}\label{d310}
s_0  = 2 \sqrt{1 - \sin\left[2(\alpha-\alpha')\right]\sin\left[2(\beta-\beta')\right]} ,
\end{align}
and
\begin{align}\label{d320}
E  & = -\cos[2(\alpha -\beta)] -  \cos[2(\alpha -\beta')] - \cos[2(\alpha' -\beta)] \nonumber \\[6pt]
&  \phantom{.x.} +  \cos[2(\alpha' -\beta')] .
\end{align}
Note that \eqref{d300} is correct because although $\hat{S}$ has four distinct eigenvalues $\pm s_0,\pm s_1$, the eigenvectors associated with $\pm s_1$ are orthogonal to $\ket{\Psi}$, so that they do not contribute  to $f_S^{\ket{\Psi}}(s)$.
Then, from \eqref{d300} it follows that
\begin{align}\label{d330}
\mean{\Psi}{\hat{S}}{\Psi} & = \int_{\mathbb{R}} s \, f_S^{\ket{\Psi}}(s) \, \di s \nonumber \\[6pt]
& = E,
\end{align}
as it should be.

So, things work fine with the operator $\hat{S}$. However, the problem is that while every physical observable of a quantum system is represented by a self-adjoint operator, the vice versa we do not know whether it is true or not \cite{coleman2020sidney}. Therefore, until someone is able to mount an experiment whose outcomes are the eigenvalues of the operator $\hat{S}$, we cannot claim that $\hat{S}$ represents a physical observable.  We remark that there has recently been a proposal to make these four measurements at once \cite{Virzì_2024} based on weak value measurements of incompatible observables \cite{PhysRevLett.117.120401,PhysRevLett.117.170402}.

\end{document}